\documentclass[english]{iopart}
\usepackage{graphicx}
\usepackage{url}
\usepackage{amssymb}
\usepackage[latin1]{inputenc}
\usepackage{epsfig}
\usepackage{iopams}

\expandafter\let\csname equation*\endcsname\relax
\expandafter\let\csname endequation*\endcsname\relax

\usepackage{amsmath}
\usepackage{txfonts}
\newcounter{fig}
\begin{document}
{\Large\hspace{.8in} Factorization of the Ising model form factors} 
\author{M. Assis$^\dag$, J-M. Maillard$^\pounds$ and B.M. McCoy$^\dag$ }
\address{$^\dag$ CN Yang Institute for Theoretical Physics, State
  University of New York, Stony Brook, NY. 11794, USA} 
\address{$^\pounds$ LPTMC, UMR 7600 CNRS, 
Universit\'e de Paris, Tour 23,
 5\`eme \'etage, case 121, 
 4 Place Jussieu, 75252 Paris Cedex 05, France} 
 
\begin{abstract}

We present a general method for analytically  factorizing 
the $n$-fold form factor integrals 
$\, f^{(n)}_{N,N}(t)$ for the correlation 
functions of the Ising model on the diagonal in terms of the  
hypergeometric functions
$\,{}_2F_1([1/2,N+1/2];[N+1];t)$ which appear in the form factor
$\,f^{(1)}_{N,N}(t)$. New  quadratic recursion 
and quartic identities are obtained 
for the form factors  for $n\,=\,\, 2,\,3$.
For $n\,=\,\, 2,\,3,\, 4$ explicit
results are given for the form factors. These 
factorizations are proved 
for all $\, N$ for $n\,=\,\, 2,\,3$.
These results yield the emergence of palindromic polynomials
canonically associated with elliptic curves.  As a consequence, 
understanding  the form factors
amounts to describing and understanding an infinite set of palindromic
 polynomials, canonically associated with elliptic curves.
From an analytical viewpoint the relation of these palindromic polynomials
with hypergeometric functions associated with  elliptic curves
is made very explicitly, and from 
a differential algebra viewpoint this corresponds to
the emergence of direct sums of differential  operators homomorphic to 
symmetric powers of a second order operator 
associated with elliptic curve.

\end{abstract}

\vskip .5cm


\noindent {\bf AMS Classification scheme numbers}: 34M55, 
47E05, 81Qxx, 32G34, 34Lxx, 34Mxx, 14Kxx 
\vskip .5cm


\section{Introduction}
\label{intro}

The form factor expansion of Ising model correlation functions  is
essential for the study of the long distance behavior and the scaling
limit of the model. This study was initiated in 1966 when Wu~\cite{wu} 
computed the first term in the expansion of the row correlations  
both for  $T> T_c$, where the result is a one dimensional integral, and for
$T<T_c$, where the result is a 2 dimensional integral. By at least
1973 it was recognized~\cite{book} that the diagonal correlations and
form factors are a specialization of the results for the row
correlations. The extension to form factors for correlations in a
general position and from the leading term to all terms
was first made in 1976~\cite{wmtb}. This leads to the general result that
for the two dimensional Ising model with interaction energy
${\mathcal E}\, \,= \,  \,\,-\sum_{j,k}\{E^v \, 
 \sigma_{j,k}\sigma_{j+1,k}+E^h\, \sigma_{j,k}\sigma_{j,k+1}$\},
with $\sigma_{j,k}\,=\,\,\pm 1$,
the form factor expansion for $T<T_c$ is
\begin{align}
\langle \sigma_{0,0}\sigma_{M,N}\rangle\, \,= \,\,\,\,  (1-t)^{1/4} \cdot
 \{1\, +\, \sum_{n=1}^{\infty} \, f^{(2n)}_{M,N}\}, 
\label{ffm}
\end{align}
where $t\,=\,\, (\sinh 2E^v/k_BT \,\sinh 2E^h/k_BT)^{-2}$, 
and for $T\, > \, T_c$
\begin{align}
\langle\sigma_{0,0}\sigma_{M,N}\rangle\,\, = \,\, \, \, (1-t)^{1/4} \cdot
 \sum_{n=0}^{\infty} \, f^{(2n+1)}_{M,N}, 
\label{ffp}
\end{align}
where $t\,=\,\, (\sinh 2E^v/k_BT \,\sinh 2E^h/k_BT)^{2}$, 
and where $f^{(n)}_{M,N}$ are $n$-fold integrals.

The form factor expansions (\ref{ffm}) and (\ref{ffp}) are of great
importance for the study of the magnetic susceptibility of the Ising model
\begin{equation}
 \chi(T) \,= \,\, \, \frac{1}{k_BT} \cdot  \sum_{M,N}\{\langle
\sigma_{0,0}\sigma_{M,N}\rangle -{\mathcal M}^2\}, 
\end{equation}
where ${\mathcal M}\,=\,\,(1-t)^{1/8}$ for $T<\,T_c$ and equals zero for $T>\, T_c$ is
the spontaneous magnetization. The study of this susceptibility
has been the outstanding problem in the field for almost 60 years. The
susceptibility is expressed in terms of the form factor expansion as
\begin{equation}
k_BT \cdot \chi(T) \, = \, \,\,  (1-t)^{1/4} \cdot \sum_{m}\chi^{(m)}(T), 
\end{equation}
where 
\begin{equation}
\chi^{(m)}(T) \, = \, \, \,  \sum_{M,N}\, f^{(m)}_{M,N}, 
\end{equation}
with $m=2n$, for $T< T_c$, and $m= \, 2n+1$, for $T> T_c$.  
In the last twelve years
a large number of remarkable properties have been obtained for both
$\chi^{(n)}(T)$~\cite{nic1}-\cite{perk} 
and the specialization to the diagonal~\cite{mccoy3}
\begin{equation}
\chi^{(n)}_d(t) \, = \, \,\,  \sum_{N,N}\, f^{(m)}_{N,N}.
\end{equation}
These remarkable properties of $\chi^{(n)}$ and $\chi^{(n)}_d(t)$ must
originate in properties of the $f^{(n)}_{M,N}$ themselves.

For 40 years after the first computations of Wu, the form factor
integrals for $n\geq 2$ appeared to be intractable in the sense that
they could not be expressed in terms  of previously known special
functions. However, in 2007 this intractability was shown to be false
when Boukraa et al~\cite{mccoy1} discovered by means of differential
algebra computations on Maple, using the form for the
form factors proven in \cite{mccoy2},  many examples for $n$ as large as nine 
that the form factors in the isotropic case $E^h=\, E^v$ 
 can be written as sums of products of the complete elliptic integrals
$K(t^{1/2})$ and $E(t^{1/2})$ with polynomial coefficients, where
 for the diagonal case ($M=\, N$) we may allow $E^v\, \neq\,  E^h$.

These computer derived examples lead to the obvious 

\vspace{.1in}

{\bf Conjecture 1}

All $n$-fold form factor integrals for Ising correlations may be
expressed in terms of sums of products of one dimensional
 integrals with polynomial coefficients. 

\vspace{,1in}

The first discovery that the $n$-fold multiple integrals which arise 
in the study of integrable models can be decomposed into sums 
of products of one dimensional integrals (or sums) was made for  
the correlation functions of the XXZ
spin chain
\begin{equation}
\label{hxxz}
 H_{XXZ}\,\,=\,\,\,\, -\sum_{j=-\infty}^{\infty}\{\sigma_j^x\sigma_{j+1}^x
\, +\sigma_j^y\sigma_{j+1}^y+\Delta \sigma_j^z\sigma_{j+1}^z\}. 
\end{equation}
These correlations were expressed as multiple integrals for the 
massive regime
($\Delta<-1$) in 1992~\cite{jmmn} and in the massless regime ($-1\leq
\Delta \leq 1$) in 1996~\cite{jm}.
In 2001 Boos and Korepin~\cite{bk} discovered that for the
case $\Delta\,=\,\, -1$, the special correlation function (called the
emptiness probability)  
\begin{equation}
P(n)\,=\,\,\,\,
\langle \prod_{j=1}^n\left(\frac{1+\sigma^z_j}{2}\right)\rangle, 
\end{equation}
for $n=\,4$ could be expressed in terms of
$\zeta(3), \,\zeta(5), \,\zeta^2(3)$ and $\ln 2$, and 
this decomposition in terms of sums of products of zeta functions of
odd argument was extended to $P(5)$ in~\cite{bkns} and $P(6)$ in~\cite{sst}.
Similar decompositions of the correlation function
 $\, \langle \sigma_0^z \sigma_n^z\rangle$
were obtained for $n=\,3$ in \cite{ssnt}, for $n=\,4$  in~\cite{bst}
and for $n=\,5$ in~\cite{ss}. The extension to the XXZ model chain 
(\ref{hxxz}) with $\Delta\,\neq \, -1$ of the decomposition of 
the integrals for the
third neighbor correlation $\langle \sigma_0^i\sigma_3^i\rangle$
 for $i=\,x,z$ 
 was made in~\cite{ksts}.

The discovery in \cite{mccoy1} that a similar reduction takes place
for Ising correlations thus leads to the more far reaching  

\vspace{.1in}

{\bf Conjecture 2}

All multiple integral representations of correlations and form factors
in all integrable models can be reduced to sums of products of one
dimensional integrals.

\vspace{.1in}

If correct this conjecture must rest upon a very deep and universal
property of integrable models.

In~\cite{mccoy1} the form factors were reduced to sums of products of
the complete elliptic integrals $K(t^{1/2})$ and $E(t^{1/2})$. However, the
results become much more simple and elegant
 when expressed in terms of the
hypergeometric functions $F_N$ and $F_{N+1}$ where
\begin{equation}
 F_N\,\, =\, \,\,\, \,\,
 _2F_1([1/2,\, N +1/2]; \, [N+1];\, t)
\label{fndef}
\end{equation}
appears in the form factor for $n=1$
\begin{align}
f^{(1)}_{N,N}(t)\,  =\,\, 
 \frac{t^{N/2}}{\pi}\,
 \cdot 
\int_0^1 \, x^{N-1/2}(1-x)^{-1/2}(1-tx)^{-1/2} \cdot dx
 \,\,=\, \, \lambda_N \cdot t^{N/2} \cdot  F_N,
\label{f1}
\end{align}
where
\begin{equation}
\lambda_N\, =\, \, \frac{(1/2)_N}{N!}, 
\end{equation}
and $(a)_0=\, 1$ and for $n\geq 1~~(a)_n=\, a(a+1)\cdots (a+n-1)$ is
Pochhammer's symbol. 
Note that $F_0 = \, \frac{2}{\pi} K(t^{1/2})  =  \,f^{(1)}_{0,0}(t)$.

The expressions for $f^{(n)}_{N,N}(t)$ in terms of $F_N$ and $F_{N+1}$ 
are obtained from~\cite{mccoy1}, rewritten by use of the contiguous relations 
for hypergeometric functions, and we
give some of these expressions in  \ref{ffbasis}. 
In all cases studied  the form factors have the form
\begin{align}
&& f^{(2n)}_{N,N}(t)\, \, =\,\, \,  \, 
 \sum_{m=0}^{n-1}\, K_m^{(2n)}\cdot  f^{(2m)}_{N,N}(t)\, 
 +\sum_{m=0}^{2n}\,  C^{(2n)}_m(N;t) \cdot  F_N^{2n-m} \cdot  F_{N+1}^{m},   
\label{feform}\\
&&\frac{f^{(2n+1)}_{N,N}(t)}{t^{N/2}}\, \, 
=\,\,\, \sum_{m=0}^{n-1}\, K_m^{(2n+1)} \cdot  
\frac{f^{(2m+1)}_{N,N}(t)}{t^{N/2}} \,\,
+\sum_{m=0}^{2n+1}\, C^{(2n+1)}_m(N;t) \cdot  F_N^{2n+1-m} \cdot F_{N+1}^m,
\label{foform}
\end{align}
where $f^{(0)}_{N,N}\,\,=\,\,\,1$. 
The degrees of the polynomials $\, C^{(j)}_m(N;t)$ 
are for $N\geq 1$
\begin{align}
{\rm deg}~C^{(2n)}_m(N;t)\,\,=\,\,\,{\rm deg}~
C^{(2n+1)}_m(N;t)\,\,\,=\,\,\,\,   n \cdot (2N+1), 
\end{align} 
with $\, C^{(n)}_m(N;t)\,\,\, \sim \,\, \, \,\,t^m\,$ as $\, t\,\sim \, 0$.

These polynomials are different from the corresponding polynomials in
the $K, \,E$ basis in that they have the palindromic property
\begin{align}
\label{pale}
C^{(2n)}_m(N;t)\,\,&= \,\,\, \, \,   t^{n(2N+1)+m} \cdot \,  C^{(2n)}_m(N;1/t),
 \\
\label{palo}
C^{(2n+1)}_m(N;t)\, \,&= \,\,\,\, \,  t^{n(2N+1)+m} \cdot \,  C^{(2n+1)}_m(N;1/t).
\end{align}

We conjecture that these results are true generally.
 
In this paper we begin the analytic proof of  Conjecture 1 and the
derivation and generalization of the results of \cite{mccoy1} for the
diagonal correlation $M=N$ by studying the three lowest order integrals 
$f^{(n)}_{N,N}(t)$ for $n=2,3,4$. The results are summarized in Sec. 2.

In Sec. 3 we derive the results for $f^{(2)}_{N,N}(t)$. We proceed by
first differentiating the integral $f^{(2)}_{N,N}(t)$ with respect to $t$, which
removes the term proportional to $f^{(0)}_{N,N}(t)$ from the general form 
(\ref{feform}). The resulting two dimensional integral is then  
seen to factorize into a sum of products of one dimensional integrals.
This factorized result is then compared with the derivative of
(\ref{feform}) to give three coupled first order inhomogeneous equations
for the three polynomials $C_m^{(2)}(N;t)$. These equations are
decoupled to give inhomogeneous equations of degree three which are
explicitly solved to find the unique polynomial solutions
$C_m^{(2)}(N;t)$.

In Sec. 4 we extend this method to $f^{(3)}_{N,N}(t)$. The first step
is to apply to $f^{(3)}_{N,N}(t)$ the second order operator which
annihilates $f^{(1)}_{N,N}(t)$. However, in this case we have not
found the mechanism which factorizes the resulting three dimensional
integral. Instead we use the property discovered in~\cite{mccoy1} that
the resulting integral satisfies a fourth order homogeneous equation
which is homomorphic to the symmetric cube of a second order operator
and thus a factorized form is obtained. This form is then compared
with the form obtained by applying the second order
operator to the form (\ref{foform}), and from this comparison we obtain
4 coupled inhomogeneous equations for the 4 polynomials
$C_m^{(3)}(t)$. These equations are then decoupled to give
inhomogeneous equations of degree 5 for $C_3^{(3)}(N;t)$ and of degree
8 for the three remaining polynomials. We then solve these equations
under the assumption that a polynomial solution exists. 

The results for  $f^{(n)}_{N,N}(t)$ with $n=1,2,3$
have a great deal of structure which can be generalized to
arbitrary arbitrary $n$. Of particular interest is the fact   
that $f^{(2n)}_{N,N}(t)$ vanishes as $t^{n(N+n)}$ and 
$f^{(2n+1)}_{N,N}(t)/t^{N/2}$ vanishes as  
$t^{n(N+n+1)}$ at $t\rightarrow 0$ while each  individual term in the
expansions (\ref{feform}) and (\ref{foform}) vanishes with a power
(which may be zero) which is independent of $N$. This cancellation
for $f^{(2)}_{N,N}(t)$ and $f^{(3)}_{N,N}(t)$ is
demonstrated in Sec. 5 and gives an interpretation of several features
of the results obtained in Secs. 3 and 4. It also provides an
alternative form (\ref{f3alt}) for $f^{(3)}_{N,N}(t)$ compared to the form
(\ref{foform}).
In Sec. 6, in a differential algebra viewpoint,
 the canonical link between the 20-th order ODEs associated with the 
 $C_m^{(4)}(N;t)$ of $f^{(4)}_{N,N}(t)$
and the theory of elliptic curves 
is made very explicit with the emergence of direct sums of differential 
operators homomorphic to symmetric powers of a second order operator 
associated with elliptic curves, and in an analytical viewpoint, 
is made very explicit with exact expressions (given in Appendix G), 
for the polynomials $C^{(4))}_m(N;t)$ , {\em valid for any $N$}. We
conclude in Sec. 7 with a discussion of possible generalizations 
of our results.

\section{Summary of formalism  and results}
\label{summary}

The form factor integrals for the diagonal
 correlations are~\cite{mccoy1,mccoy2}
for $T\,<\,T_c$
\begin{align}
f^{(2n)}_{N,N}(t)\, \,\,  &=\,\,\,  \frac{t^{n(N+n)}}{(n!)^2\,\pi^{2n}}\,
\int_0^1\, \prod_{k=1}^{2n}\,dx_k\,\, x_k^N\,\prod_{j=1}^n\,
\left(\frac{(1-tx_{2j})(x_{2j}^{-1}-1)}{(1-tx_{2j-1})
(x_{2j-1}^{-1}-1)}\right)^{1/2}
\nonumber\\
&\qquad\prod_{1\leq j \leq n}\,\prod_{1\leq k \leq n}
\,\left(\frac{1}{1-tx_{2k-1}x_{2j}}\right)^2
\,\prod_{1\leq j<k \leq n}(x_{2j-1}-x_{2k-1})^2\,(x_{2j}-x_{2k})^2,
\nonumber\\
\label{dffm}
\end{align}
and for $T\, > \, T_c$
\begin{align}
f^{(2n+1)}_{N,N}(t) &  = \nonumber\\
&\frac{t^{(n+1/2)N+n(n+1)}}{n!(n+1)!\pi^{2n+1}}
\,\int_0^1\,\prod_{k=1}^{2n+1}\,dx_k\,\,x_k^N
\prod_{j=1}^{n+1}\,x_{2j-1}^{-1}[(1-tx_{2j-1})(x^{-1}_{2j-1}-1)]^{-1/2}
\nonumber\\
&\prod_{j=1}^{n}\,x_{2j}[(1-tx_{2j})\,
(x^{-1}_{2j}-1)]^{1/2}\, \prod_{1\leq j \leq n+1}\, \prod_{1\leq k \leq n}
\left(\frac{1}{1-tx_{2j-1}x_{2k}}\right)^2
\nonumber\\
&\prod_{1\leq j <k\leq n+1}\, (x_{2j-1}-x_{2k-1})^2 \, 
\prod_{1\leq j <k \leq n}\,(x_{2j}-x_{2k})^2.
\label{dffp}
\end{align}

When $t=0$ the integrals in (\ref{dffm}) and (\ref{dffp}) 
reduce to a special case of the Selberg integral~\cite{sel,wf}
\begin{align}
f^{(2n)}_{N,N}(t)\sim \, \, &\, \,  \frac{t^{n(N+n)}}{(n!)^2\pi^{2n}}
\frac{\Gamma(N+n+1/2)\Gamma(n+1/2)}{\Gamma(N+1/2)\Gamma(1/2)}
 \nonumber\\
&\times \, \prod_{j=0}^{n-1}\left[\frac{\Gamma(N+j+1/2)\Gamma(j+1/2)\Gamma(j+2)}
{\Gamma(N+n+j+1)}\right]^2
\end{align}
and 
\begin{align}
f^{(2n+1)}_{N,N}(t)\sim \, & \, \frac{t^{N(n+1/2)+n(n+1)}}{n! \pi^{2n+1}}
\frac{\Gamma(N+1/2)\Gamma(1/2)}{\Gamma(N+n+1)} \nonumber\\
&\times \, \prod_{j=0}^{n-1}\left[\frac{\Gamma(N+j+3/2)\Gamma(j+3/2)\Gamma(j+2)}
{\Gamma(N+n+j+2)}\right]^2
\end{align}
In particular
\begin{align}
&f^{(2)}_{N,N}(t)\, =\,\,\,
  t^{N+1} \cdot \frac{\lambda_{N+1}^2}{(2N+1)}\,\,+O(t^{N+2}), 
\label{f2t0} \\
&f^{(3)}_{N,N}(t)\,
=\,\,\,t^{3N/2+2} \cdot \frac{\lambda_{N+1}^3}{2(2N+1)(N+2)^2}\,\,
+O(t^{3N/2+3}).
 \label{f3t0}
\end{align}

\subsection{General Formalism}
\label{General}

For the special case $f^{(2)}_{N,N}(t)$ we will analytically derive
the form (\ref{feform}) without making any
assumptions. However, for the general case we will proceed by
assuming the forms (\ref{feform}) and (\ref{foform}) as an ansatz and
with this as a conjecture, we will derive inhomogeneous Fuchsian
equations for the polynomials $C^{(n)}_m(N;t)$
\begin{equation}
\label{OCI}
\Omega^{(n)}_m(N;t)\cdot C^{(n)}_m(N;t)\,\,\,  =\,\,\,\, \,  I^{(n)}_m(N;t),
\end{equation}
where $\Omega^{(n)}_m(N;t)$ is a linear 
differential operator and $I^{(n)}_m(N;t)$ a polynomial. 

In all cases which have been studied, the operator 
$\, \Omega^{(n)}_m(N;t)$, corresponding to  the lhs of (\ref{OCI}),
has a {\em direct sum decomposition} where each term in the direct sum is
{\em homomorphic to a either a symmetric power
or a symmetric product
for different values of $\, N$, 
of the second order operator}
\begin{equation}
\label{o3}
O_2(N;t)\,\, = \, \, \,\,\,  D_t^2\,\, \,  -\frac{1+N-Nt}{t(1-t)} \cdot D_t \, \, 
+\frac{4 \,+4N\, -t \,-2Nt}{4\, t^2\,(1-t)}, 
\end{equation}
where $D_t\, = \, \, d/dt$.  
The operator $O_2(N;t)$ is equivalent to the operator
 $L_2(N;t)$ which annihilates
$f^{(1)}_{N,N}(t)$ \cite{mccoy1}, as can be seen in the operator isomorphism
\begin{align}
\label{O2L2}
O_2(N;t) \cdot t^{N/2\,+\,1} \, \,\,  \,=\,\,\, \, \, \,  t^{N/2\,+\,1} \cdot L_2(N;t).
\end{align}

The solutions of $\,O_2(N;t)$  are expressed in terms
of  hypergeometric functions by noting that
\begin{align}
t^2 \cdot (1-t) \cdot O_2(N) 
\,\, =  \,\, t \cdot  (t \, D_t\, +a)(t \, D_t\, +b)
\,\,  -(t\, D_t\, -a')(t\, D_t\, -b'), 
\nonumber 
\end{align}
with 
\begin{align}
\label{abapbp}
a\, =\, -N-1/2, \quad \,  \,  \, b\, =\, \, -1/2, \quad \,  \,   \,
a'\, =\,\,  N\, +1, \quad \,  \,  \, b'\, =\,\,  1, 
\end{align}
which for $|t|<1$
\footnote[2]{For $|t|> 1$, we write $z=\,1/t$ and the 
identical procedure is found to
interchange $a$ with $a'$ and $b$ with $b'$. Thus
the two fundamental solutions valid near $t=\infty$ are
$\,{\tilde u}_1(N;z)\,\, =\,\, z^{-1/2}\cdot \, _2F_1([1/2,1/2+N];[1+N];z)=\,z^{-1/2}\cdot F_N$,
$     \, {\tilde u}_2(N;z)\,\, =\,\, z^{-N-1/2} \cdot \, _2F_1([1/2,1/2-N];[1-N];z)$.
The identification of the hypergeometric functions of (\ref{u1}) and
(\ref{u2}) with these two solutions is a consequence
of the palindromic property of the operator $O_2(N;t)$. However, we 
note that ${\tilde u}_j(N;z)$ is not the analytic continuation of
$u_j(N;t)$.}
has the two fundamental solutions~\cite[p. 283]{ww}
\begin{align}
t^{a'} \cdot \, \, _2F_1([a+a',b+a']; \, [a'-b'+1]; \, t), \,\,
 t^{b'} \cdot {}_2F_1([a+b',b+b']; \, [b'-a'+1]; \, t). 
\end{align}
Using (\ref{abapbp}) we have the two solutions of $O_2(N)$
\begin{align}
\label{u1}
&u_1(N;t)\,\,  =\,\, \,  \, 
 t^{N+1} \cdot \, _2F_1([1/2,1/2+N]; \, [N+1]; \, t) 
\,\, =\,\, \,  t^{N+1} \cdot \, F_N,  \\
\label{u2}
&\hbox{and:} \qquad \quad ~\,    t \cdot \, _2F_1([1/2,\, 1/2-N]; \, [1-N]; \,t).
\end{align} 

The solution $u_1(N;t)$ in (\ref{u1}) is regular at $t=\, 0$
 and has the expansion
\begin{equation}
\label{u1def}
u_1(N;t) \, \,=\quad  t^{N+1} \cdot 
\sum_{n=0}^{\infty}\,b_n(N)  \cdot t^n,
\end{equation}
with
\begin{equation}
b_n(N)\,\, =\,\, \,\frac{(1/2)_n \, (1/2+N)_n}{(N+1)_n\, n!}.
\label{bndef}
\end{equation}

Since we will in this paper
work with positive integer values of $\, N$,
it is better to introduce as the second solution
\begin{align}
\label{second}
t^{N+1} \cdot \, _2F_1([1/2,\, 1/2+N]; \, [1]; \,1\, -\, t).
\end{align}
When $\, N$ is {\em not an integer} 
the hypergeometric function (\ref{second})
 can be written as the following linear combination
of the two previous solutions (\ref{u1}) and (\ref{u2})
\begin{align}
&{{\Gamma(-N)} \over {\Gamma(1/2)\, \Gamma(1/2-N)}} \cdot 
 t^{N+1} \cdot \, _2F_1([1/2,1/2+N]; \, [N+1]; \, t) \nonumber \\
& + \, \, \, 
{{\Gamma(N)} \over {\Gamma(1/2)\, \Gamma(1/2+N)}} \cdot 
 t \cdot \, _2F_1([1/2,\, 1/2-N]; \, [1-N]; \,t).
\end{align} 

The hypergeometric function (\ref{second})
 is not analytic at $t=\, 0$ but, instead, has a
{\em logarithmic singularity}.

 From~\cite[(2) on p.74 and (7) on p.75]{bateman}
we may choose to normalize the analytical part of the
second solution to $t$ as $t\rightarrow 0$. 
Denoting such a solution  $u_2(N;t)$, it reads
\begin{align}
\label{u2def}
u_2(N;t) \, \, = \, \, \, \, 
t \cdot \sum_{n=0}^{N-1} \, a_n(N) \cdot t^{n}
+t^{N+1} \cdot N \cdot \lambda_N^2  \cdot 
\sum_{n=0}^{\infty} \, b_n(N) \, 
[k_n-\ln(t)] \cdot t^{n},
\end{align}
with $a_0(N) \, \, =\, \, 1$ and for $n\geq 1$
\begin{equation}
a_n(N) \, \, =\, \, \, \frac{(1/2)_n\,(1/2-N)_n}{(1-N)_n \, n!}
 \, = \, \,
\lambda_N \cdot \frac{(1/2)_n\,(N-n)!}{(1/2)_{N-n}\,n!}
\label{andef}
\end{equation}
and $\, k_n \,= \, H_n(1)+H_{n+N}(1)-H_n(1/2)-H_{n+N}(1/2)$, 
where 
\begin{equation}
\label{hndef}
H_n(z) \, \, = \,  \, \, \sum_{k=0}^{n-1} \, \frac{1}{z+k}
\end{equation}
are the partial sums of the harmonic series.
The series expansion (\ref{u2def})
 corresponds to the maximal unipotent
 monodromy structure of $\,O_2(N;t)$
which amounts to writing the second solution as: 
\begin{align}
\label{u2defbis}
u_2(N;t) \, \, = \, \, \, \, w_2(N; t)\, \, \,
- \, N \cdot \lambda_N^2 \cdot u_1(N;t) \cdot \ln(t)  
\end{align}
where $\, w_2(N; t)\, = \, \,t \,\, + \, \cdots \,  \,   
$ is analytical at $t \, = \, 0$.
This  function  $\, w_2(N; t)$
is the solution analytic at $\, t\, = 0$, different from  $u_1(N;t)$, of
 an order-four operator  which factorizes
as the product $\, \tilde{O}_2(N;t) \cdot O_2(N;t)$,  where $\, \tilde{O}_2(N;t)$ and $\, O_2(N;t)$
are  two order-two homomorphic operators 
\begin{align}
\tilde{O}_2(N;t) \cdot I_1 \, \, \, = \, \, \, \, 
J_1 \cdot \, O_2(N;t),
\end{align}
where one of the two order-one intertwinners $\, I_1$ and $\, J_1$ is quite simple, namely
\begin{align}
I_1 \, \, = \, \, \, \, {{1} \over {t}}  \cdot D_t  \, \, - {{t\, -2} \over {2\, t^2 \cdot (t\, -1)}}
 \, \, -\, {{N} \over {2\, t^2}}. 
\end{align}

Finally, we note the relation
which follows from the Wronskian of $ \, O_2(N;t)$,
\begin{equation}
\label{u1u2}
u_1(N) \cdot u_2(N+1)\, -\beta_N \cdot u_2(N) \cdot u_1(N+1)\,\, =\,\,\,  t^{N+2}, 
\end{equation}
with
\begin{equation}
\label{abdef}
\beta_N\, =\, \,\, \frac{(2N+1)^2}{4N(N+1)}.
\end{equation}

\subsection{Explicit results for $f^{(2)}_{N,N}(t)$}
\label{subf2} 

For $f^{(2)}_{N,N}(t)$ the parameter $K^{(2)}_0$ and the polynomials
$C^{(2)}_m(N;t)$ of the form (\ref{feform}) 
are explicitly computed in Section \ref{case} as
\begin{equation}
\label{cdef}
K^{(2)}_0 \, \, = \, \, \, N/2, 
\end{equation}
and
\begin{equation}
\label{initial}
C^{(2)}_m(N;t)\,\, =\,\,\, \, \,
A^{(2)}_{m}\cdot t^m \cdot \sum_{n=0}^{2N+1-m}\,c^{(2)}_{m;n}(N) \cdot t^n,
\end{equation}
with
\begin{align}
A^{(2)}_n \, = \,  \,  \, (-1)^{n+1} \cdot  \frac{N}{2} \cdot  {n \choose 2} \cdot \beta^n_N.
\label{final}
\end{align}
Using the notation that
\begin{equation}
[f]_n=\, =\, \,  {\rm the~coefficient~ of}~t^n~{\rm in~the~expansion~of}~f~{\rm
  at}~t=0
\end{equation}
 we have for $ \, 0 \, \leq \,  n \, \leq \,  N-1$
\begin{align}
\label{help0}
c^{(2)}_{2;n}(N)\,\,  &=\,\, \, \, c^{(2)}_{2;\, 2N-1-n}(N)
=\,\, [t^{-2}u_2(N)^2]_n
\,\,  \, = \, \,\,\, \sum_{k=0}^{n}\, a_k(N)  \cdot a_{n-k}(N), \\
\label{dn1}
c^{(2)}_{1;n}(N)\, \,  &=\, \,  c^{(2)}_{1;\, 2N-n}(N)
=\,\,  [t^{-2}u_2(N)u_2(N+1)]_n\, \,   =\,\, \,\,
 \sum_{k=0}^{n}\, a_k(N) \cdot  a_{n-k}(N+1), 
\end{align}
and
\begin{equation}
\label{c21n}
c^{(2)}_{1;N}(N)\,\, =\, \, \lambda_N^2\,\,  +c^{(2)}_{2,N-1}(N), 
\end{equation}
and where for $ \, 0 \, \leq \,  n \, \leq \,  N$
\begin{equation}
c^{(2)}_{0;\, n}(N) \,\, =  \,\,\,c^{(2)}_{0;\, 2N+1-n}(N) \,\, =\, \, 
[t^{-2}u_2^2(N+1)]_n \,\,\,=\,\,   c^{(2)}_{2;\, n}(N+1),
\label{c2n0} 
\end{equation}
where $a_n(N)$ is given by (\ref{andef}).
We note that the sum (\ref{help0}) for $c^{(2)}_{2,N-1}$ may be written
by use of the second form of $a_{n}(N)$ in (\ref{andef}) 
in the alternative form 
\begin{equation}
c^{(2)}_{2;N-1}\,= \,\, \, \lambda_N^2 \cdot \, 2N \cdot H_N(1/2),  
\end{equation}
where $\, H_N(z)$ is given by (\ref{hndef}).

We also derive the recursion relation for $N\, \geq\,  1$
\begin{equation}
\label{nice}
f^{(2)}_{N,N}(t)\,\, =\,\, \,\, \,  N\,   f^{(2)}_{1,1}(t)\, \,  \,\,  
- \frac{N}{2}\,\, t^{1/2} \cdot 
\sum_{j=1}^{N-1}\, \frac{f^{(1)}_{j,j}(t)\cdot  f^{(1)}_{j+1,j+1}(t)}{j(j+1)}.
\end{equation}

\subsection{Explicit results for $f^{(3)}_{N,N}(t)$}
\label{explicit}

For $f^{(3)}_{N,N}(t)$ the parameter $K^{(3)}_0$ and the polynomials
$C^{(3)}_m(N;t)$ of the form (\ref{foform}) 
are explicitly computed in Section \ref{ffactor} as
\begin{align}
K^{(3)}_0\,\, &=\,\,\, \frac{3\, N+1}{6},\label{k30} 
\end{align}
and
\begin{equation}
\label{c02form}
C^{(3)}_m(N;t)\,\, =\,\,\, \, \,
A^{(3)}_{m}\cdot t^m \cdot \sum_{n=0}^{2N+1-m}\,c^{(3)}_{m,n}(N) \cdot t^n \, \, 
+ \frac{N-1}{N}\lambda_N\cdot C^{(2)}_m(N,t),
\end{equation}
where we make the definition $C^{(2)}_3(N,t)=0$ and 
\begin{equation}
A^{(3)}_n\, =\, \,\, 
 (-1)^{n+1}\cdot  \frac{2}{3} \cdot \, {n \choose 3}\,\cdot \lambda_N \cdot \beta^n_N.
\label{a3n}
\end{equation}
The coefficients
$c^{(3)}_{m;n}(N)$'s are given by a simple {\em quartic expression}
 of the $a_n$'s and  $b_n$'s.
For $0\leq n \leq N-1$ they read
\begin{align}
c^{(3)}_{3;n}(N)\, \, \, &=\,\, \,\,   \,
 c^{(3)}_{3;\, 2N-2-n}(N)\, \,  \,= \,\,\, [t^{-N-4}u_2^3(N)u_1(N)]_n
\,\, =  
\nonumber\\
\quad \,  \,&=\,\,  \,  \,
\sum_{m=0}^n\,\sum_{l=0}^m\sum_{k=0}^l \, a_k(N) \cdot
a_{l-k}(N)\cdot a_{m-l}(N)\cdot b_{n-m}(N), \label{c33n} 
\end{align}
and
\begin{align}
c^{(3)}_{2;n}(N)\, \,&=\,\, c^{(3)}_{2;\, 2N-n-1}(N)\,\,
=\,\, \,[t^{-N-4}u_2^2(N)u_2(N+1)u_1(N)]_n
\,\,=  \nonumber \\
 \quad \, \,&=\,\,\, \, \sum_{m=0}^n\,\sum_{l=0}^m\,\sum_{k=0}^l \, \, 
a_k(N) \cdot a_{l-k}(N) \cdot  a_{m-l}(N+1) \cdot  b_{n-m}(N), 
\label{c32n}
\end{align}
for $0 \,\leq\, n \leq \, N$
\begin{align}
c^{(3)}_{0;n}(N)\,\,  &=\,\,\,c^{(3)}_{0;\, 2N-n+1}(N)\,\,
\,\, =\,\, \, [t^{-N-4}u_2^3(N+1)u_1(N)]_n 
\nonumber\\
\quad \, \,&=\,\,\, 
\sum_{m=0}^n\,\sum_{l=0}^m\,\sum_{k=0}^l\, a_k(N+1)
\cdot  a_{l-k}(N+1) \cdot  a_{m-l}(N+1) \cdot  b_{n-m}(N),
\label{c30n}
\end{align}
and for $0 \,\leq\, n\,\leq \,N-1$
\begin{align}
c^{(3)}_{1;n}(N)\,\, &=\,\,\, \,  c^{(3)}_{1;\, 2N-n}(N)\, \,=\,\,\,
[t^{-N-4}u_2(N)u_2^2(N+1)u_1(N)]_n
  \nonumber\\
 \quad \, \,\,&=\,\,\, \sum_{m=0}^n\sum_{l=0}^m \sum_{k=0}^l\,
a_k(N) \cdot a_{l-k}(N+1)\cdot a_{k-m}(N+1)\cdot b_{n-m}(N), 
\label{c31n}
\end{align}
with the middle term of $C^{(3)}_1(N;t)$ of order $N+1$
\begin{align}
c^{(3)}_{1;\, N} \, &= \,\, \frac{\beta_N \cdot \lambda_N^2}{N}\, \, \,
 +\beta_N \cdot \lambda_N \cdot  
\left[(N-1)\cdot c^{(3)}_{2;\, N-1} \,+4\,c^{(2)}_{2;\, N-1}\right]
 \nonumber\\ 
&\quad - \frac{2}{3} \cdot \frac{\beta_N \cdot \lambda_N}{N^2}\cdot 
\left[2N^2 \cdot c^{(3)}_{3;\, N-2} \, +(N^2-1/4) \cdot c^{(3)}_{3;\, N-1}\right] ,
   \label{c31middle}
\end{align}
where $a_n(N)$ and $b_n(N)$ are given by (\ref{andef}) and
(\ref{bndef}).

\section{The derivation of the results for $f^{(2)}_{N,N}(t)$}
\label{case}

We begin our derivation  of the results for $f^{(2)}_{N,N}$ 
of Sec. \ref{subf2}   by integrating (\ref{dffm}) (with $2n=2$)  by parts
using  
\begin{align}
 \label{part5}
u \, \, &=  \, \,\,  y^{N-1/2}\cdot (1-y)^{1/2}\cdot (1-ty)^{1/2}, 
 \\
 \label{part6}
du \, \, &= \, \,\,  y^{N-3/2}
\frac{[N \cdot  (1-y)\cdot (1-ty)\,
 -1/2\, (1\, -t\,y^2)]}{(1-y)^{1/2} \cdot (1\,-t\,y)^{1/2}} \cdot  dy,  
\\
\label{part8}
dv \, \,  &=  \,\, \,   \frac{dy}{(1\,-t\,xy)^2}, \quad \qquad 
v \,\,   = \, \, \,  \frac{y}{1-txy}, 
\end{align}
to find
\begin{align}
\label{partialint1}
f^{(2)}_{N,N}(t) \,\,  &=   \,\, \,
\nonumber \\
&\int_0^1\!{dx}\, \int_0^1\!{dy\ \frac{t^{N+1}}{2\pi^2}
\frac{x^{N+1/2}\, y^{N-1/2}\, (1-ty^2)}
{(1-x)^{1/2}\, (1\,-t\,x)^{1/2}\, (1-y)^{1/2}\, (1\,-t\,y)^{1/2}\, (1-txy)}} 
\nonumber \\
& \quad - N\, \int_0^1\!{dx}\, \int_0^1\!
{dy\ \frac{t^{N+1}}{\pi^2}\ \, 
  \frac{x^{N+1/2}\, y^{N-1/2}\, (1-y)^{1/2}\, (1-ty)^{1/2}}
{(1-x)^{1/2}\, (1-tx)^{1/2}\, (1-txy)}}. 
\end{align}
The first term in (\ref{partialint1}) is separated into two parts as
\begin{align}
\int_0^1\!\, {dx}\, 
\int_0^1\!\, {dy\ \, \frac{t \cdot x^N\, y^N}{2\pi^2}\, \frac{x^{1/2}}{y^{1/2}}
\frac{1}{(1-x)^{1/2}\, (1-tx)^{1/2}\, (1-y)^{1/2}\, (1-ty)^{1/2}\, (1-txy)}}
\nonumber\\
 - \int_0^1\!{dx}\int_0^1\!{dy\ \frac{t \cdot x^N\, y^N}{2\pi^2}
\frac{t\, x^{1/2}y^{3/2}}{(1-x)^{1/2}\,
 (1-tx)^{1/2}\, (1-y)^{1/2}\, (1-ty)^{1/2}(1-txy)}},  
 \label{partialint2}
\end{align}
and in this second term we interchange $\, x\,\, \leftrightarrow\,\,  y$.
Then, recombining these two terms,
 we see that the factor $1\,-txy$ cancels
between the numerator and denominator
 in~(\ref{partialint2}). Thus the first term in
(\ref{partialint1}) factorizes and we find
\begin{align}
f^{(2)}_{N,N}(t)\, \, &= \,\,\, \int_0^1\! {dx}\int_0^1 \!{dy\
 \,  \frac{t^{N+1}}{2\pi^2}\frac{x^{N+1/2}y^{N-1/2}}
{(1-x)^{1/2}\, (1-tx)^{1/2}\, (1-y)^{1/2}\, (1-ty)^{1/2}}} 
 \nonumber \\
& \qquad - N \, \int_0^1\!{dx} \,\int_0^1\! \, {dy\ \, \frac{t^{N+1}}{\pi^2}\
  \frac{x^{N+1/2} \, y^{N-1/2} \, (1-y)^{1/2} \, (1-ty)^{1/2}}
{(1-x)^{1/2}\, (1-tx)^{1/2}\, (1-txy)}} 
\nonumber\\  
&=\,\,\,\,\, \frac{t^{1/2}}{2} \cdot f^{(1)}_{N,N} \cdot f^{(1)}_{N+1,N+1}
\nonumber\\
& \qquad - N \, \int_0^1\!{dx} \, \int_0^1\!
{dy\ \frac{t^{N+1}}{\pi^2}\  \,
  \frac{x^{N+1/2} \, y^{N-1/2} \, (1-y)^{1/2}\, (1-ty)^{1/2}}
{(1-x)^{1/2} \, (1-tx)^{1/2} \, (1-txy)}}. \label{f2nparts}
\end{align}
From (\ref{dffm}) we find for $N\,\geq\, 1$ 
that the integral in the second term of
(\ref{f2nparts}) is 
$\,f^{(2)}_{N,N}(t)\,\, -f^{(2)}_{N+1,N+1}(t)$ and thus we have
\begin{equation}
f^{(2)}_{N,N}(t)\,\,\, =\,\,\, \, \,
\frac{t^{1/2}}{2} \cdot f^{(1)}_{N,N}(t)  \cdot f^{(1)}_{N+1,N+1}(t)
\,\,
-N \cdot [f^{(2)}_{N,N}(t)\,\,-f^{(2)}_{N+1,N+1}(t)].\label{parts3} 
\end{equation}
From (\ref{parts3}) we obtain the recursion relation
\begin{equation}
f^{(2)}_{N+1,N+1}(t)\,\, =\, \,\, \, \frac{N+1}{N} \cdot f^{(2)}_{N,N}(t)
\,\,
-\frac{t^{1/2}}{2\, N} \cdot  f^{(1)}_{N.N}(t) \cdot  f^{(1)}_{N+1,N+1}(t), 
\end{equation}
and thus for $N\, \geq\,  1$
\begin{equation}
f^{(2)}_{N,N}(t)\,\, =\,\, \,\, \,  N\,   f^{(2)}_{1,1}(t)\, \,  \,\,  
-\frac{N}{2} \,   t^{1/2} \cdot 
\sum_{j=1}^{N-1}\, \frac{f^{(1)}_{j,j}(t)\cdot  f^{(1)}_{j+1,j+1}(t)}{j(j+1)}.
\end{equation}

To proceed further we return to (\ref{f2nparts}) which we write in
terms of $F_N$ as
\begin{align}
\label{f2npart3} 
f^{(2)}_{N,N}(t)\,\, &=\, \, \,\,\, 
\frac{\lambda_N \, \lambda_{N+1}}{2} \cdot t^{N+1} \cdot  F_N \cdot  F_{N+1}
\nonumber\\
\nonumber\\
&\quad \quad \, \, 
- N \, \int_0^1\!{dx}\int_0^1\!{dy\ \, \frac{t^{N+1}}{\pi^2}\
  \frac{x^{N+1/2}\, y^{N-1/2} \, (1-y)^{1/2} \, (1-ty)^{1/2}}
{(1-x)^{1/2} \,(1-tx)^{1/2}\, (1-txy)}}.
\end{align}
The integral in (\ref{f2npart3}) does not have a manifest 
factorization. However, if we compute
 $\, df^{(2)}_{N,N}(t)/dt$ in the contour
integral form of (\ref{dffm}), and note that
\begin{align}
&\frac{d}{dt}
\left[\frac{(y-t^{1/2})\, (1-t^{1/2}y)}
{(x-t^{1/2})\, (1-t^{1/2}x)}\right]^{1/2}
\, \, \, \nonumber\\
&\,\, =\,\, \frac{1}{t^{1/2}}
\left[\frac{(y-t^{1/2})(1-t^{1/2}y)}{(x-t^{1/2})(1-t^{1/2}x)}\right]^{1/2} 
\frac{(xy-1)(x-y)(t-1)}{(y-t^{1/2})(1-t^{1/2}y)(x-t^{1/2})(1-t^{1/2}x)},
\label{split}
\end{align}
the resulting integral does
factorize and, introducing  $G_N$,
 some well-suited linear combination of $F_N$ and $F_{N+1}$, 
\begin{eqnarray}
&&G_N \,\, = \, \,\,  _2F_1([3/2,N+3/2]; \, [N+1]; \, t) 
\nonumber \\
&&\qquad  \,\, = \, \,\,   {{1+t} \over {(1-t)^2}} \cdot F_N \,\,
 - \, \,  {{t} \over {(1-t)^2}} \cdot
 {{2\, N\, +1} \over { N\, +1}}  \cdot F_{N+1}, 
\end{eqnarray}
 we find
\begin{multline}
\label{df2n}
\frac{df^{(2)}_{N,N}(t)}{dt}\,\, =\, \,\, \, (1-t) \cdot t^N \cdot
\frac{(2N+1)\lambda^2_N}{16\, (N+1)}
\cdot \left[(2N+1)^2 \cdot
F_{N+1} \cdot G_N \right.
\\
 \left. -(2N-1)(2N+3)\cdot F_N \cdot G_{N+1}\right].
\end{multline}
%where $F_N$ is defined by (\ref{fndef}).
 
It remains to integrate (\ref{df2n}). However, in general, integrals
of products of two hypergeometric functions with respect 
to the argument will not have the form of the product of two 
hypergeometric functions. We will thus proceed in the opposite
direction by differentiating
(\ref{feform}) for $2n\, =\, 2$ with respect to 
$t$ and equating the result to
(\ref{df2n}) to obtain differential equations for the $C^{(2)}_m(N;t)$
which we will then solve to obtain the final results
(\ref{cdef})--(\ref{final}).

From a straightforward use of the contiguous relations of
hypergeometric functions~\cite{bateman}, we  
introduce the following well-suited linear combination of $F_N$ and $F_{N+1}$
\begin{eqnarray}
&& {\bar F}_N\,\, =\,\,\,\, \,_2F_1([3/2,N+3/2]; \, [N+2]; \, t)
\,\, =\,\,\,\, \, \frac{4\, (N+1)}{2 N+1} \cdot  \frac{dF_N}{dt}
 \nonumber\\
&&\qquad 
 \,\, =\,\,\,\, \, 
  {{1} \over{1-t}} \cdot   \Bigl(2 \cdot(N+1) \cdot  F_N 
\, \, -(2\, N+1) \cdot F_{N+1} \Bigr).
\end{eqnarray}
The derivative of
(\ref{feform}) with $2n=\, 2$
may be written in the quadratic form\footnote[3]{For convenience 
the dependence of the $C^{(2)}_m$ on $N$ and $t$ is
 suppressed here and below (see (\ref{form11})).} 
\begin{equation}
\label{form10}
B_1  \cdot F_N^2\, \, \, \,+\,B_2 \cdot F_N \cdot {\bar F}_N\, 
\,\,\,  +\,B_3 \cdot  {\bar F}_N^2, 
\end{equation}
with
\begin{align}
\label{dform1}
&B_1 \, \, = \,\,  \, \,  \, \frac{dC^{(2)}_0}{dt}\,\,
-\frac{(N\mathrm{+}1)}{2(N\mathrm{+}1/2)t} \cdot C^{(2)}_1
+\frac{(N\mathrm{+}1)}{(N\mathrm{+}1/2)}\cdot \frac{dC^{(2)}_1}{dt} \,
\nonumber\\
&\qquad 
-\frac{(N\mathrm{+}1)^2}{(N\mathrm{+}1/2)^2\, t} \cdot C^{(2)}_2\,\,  
+\frac{(N\mathrm{+}1)^2}{(N\mathrm{+}1/2)^2} \cdot 
\frac{dC^{(2)}_2}{dt}, 
\end{align}
\begin{align}
\label{dform2} 
B_2  \,  \,  & =  \,  \,   \, \,  
 \frac{(N\mathrm{+}1/2)}{(N\mathrm{+}1)} \cdot C^{(2)}_0
\nonumber \\
&\qquad
+\left[1+\frac{1}{2(N\mathrm{+}1/2)}\,
+\frac{1-2t+N(1-t)}{2(N\mathrm{+}1/2)t}\right] \cdot C^{(2)}_1 \, \, 
\, -\frac{(1-t)}{2(N\mathrm{+}1/2)} \cdot \frac{dC^{(2)}_1}{dt}\,
\nonumber \\
&\qquad 
+\frac{(N\mathrm{+}1)(3\mathrm{+}2N\mathrm{-}2t)}
{2(N\mathrm{+}1/2)^2t}\cdot C^{(2)}_2\,\, 
-\frac{(N\mathrm{+}1)(1\mathrm{-}t)}{(N\mathrm{+}1/2)^2} 
\cdot \frac{dC^{(2)}_2}{dt}, 
\end{align}
\begin{equation}
\label{dform3}
B_3  \,  = \,\,  \, -\frac{(1-t)}{4(N+1)} \cdot C^{(2)}_2
 \,  \, 
-\frac{(2\mathrm{+}2N\mathrm{-}t)(1\mathrm{-}t)}
{4(N\mathrm{+}1/2)^2t} \cdot C^{(2)}_2 \, 
 +\frac{(1\mathrm{-}t)^2}{4(N\mathrm{+}1/2)^2} 
 \cdot \frac{dC^{(2)}_2}{dt}.
\end{equation}
% with
% \begin{equation}
% \frac{dF_N}{dt} \,  \,= \,  \,\, \,
% \frac{N+1/2}{2(N+1)} \cdot {\bar F}_N.
% \end{equation}
The derivative of $f_{N,N}^{(2)}(t)$ 
in (\ref{df2n}) 
 by use of contiguous relations~\cite{bateman}
is  expressed in terms of
$F_N$ and ${\bar F}_N$ as 
% (again suppressing the dependence of $C_j$ on $N$ and $t$)
\begin{equation}
\label{form11}
\frac{df^{(2)}_{N,N}(t)}{dt}\,\, \,   =\,\, \, \, \, \,
 B_4 \cdot F_N^2\, \, \,+B_5 \cdot F_N\cdot {\bar F}_N\,\, 
 +B_6 \cdot {\bar F}_N^2, 
\end{equation}
where,
\begin{eqnarray}
&&B_4  \,\,  =  \, \, \,  \frac{2N+1}{4}   \cdot \lambda^2_Nt^N,\nonumber\\ 
&&B_5  \,\,  = \,  \, \,  \frac{[t\, -N(1-t)](2N+1)}{4(N+1)}
\cdot \lambda_N^2t^N,\nonumber\\
&& B_6  \, \, =\,   \, \, -\frac{N\beta_N\lambda_N^2 }
{4\,\, (N+1)} \cdot   (1-t) \cdot t^{N+1} 
\label{df2n2}
 \end{eqnarray}

\subsection{Linear differential equations for $C^{(2)}_m(N;t)$}
\label{diffeqC2}

To obtain the $C^{(2)}_m(N;t)$ we 
equate (\ref{form10}) with (\ref{form11}) and find the
following first order system of equations for $C^{(2)}_m(N;t)$
\begin{multline}
\label{eqn1}
\frac{(2N+1)}{4} \cdot \lambda_N^2  \cdot t^N
\, \,\,= \,\,\,\, \,\,  \frac{dC^{(2)}_0}{dt}\, \, \, 
-\frac{(N\mathrm{+}1)}{2(N\mathrm{+}1/2)t} \cdot C^{(2)}_1 \, \,
+\frac{(N\mathrm{+}1)}{(N\mathrm{+}1/2)} \cdot \frac{dC^{(2)}_1}{dt} \,  \, \, 
\\ 
\quad \quad 
-\frac{(N\mathrm{+}1)^2}{(N\mathrm{+}1/2)^2t} \cdot C^{(2)}_2 \, \, 
+\frac{(N\mathrm{+}1)^2}{(N\mathrm{+}1/2)^2} \cdot \frac{dC^{(2)}_2}{dt}, 
\end{multline}
\begin{multline}
\frac{(2N+1)\cdot [t-N(1-t)]}
{4\, (N+1)}  \cdot \lambda_N^2 \cdot t^N \,\,
= \, \,\,\, \frac{(N\mathrm{+}1/2) }{(N\mathrm{+}1)} \cdot C^{(2)}_0\\
\label{eqn2}
\quad  +\left[1\,+\frac{1}{2(N\mathrm{+}1/2)} \,
+\frac{1-2t\,+N\,(1-t)}{2(N\mathrm{+}1/2)t}\right] \cdot C^{(2)}_1 \,\,
-\frac{(1-t)}{2\,(N\mathrm{+}1/2)}\frac{dC^{(2)}_2}{dt} \\
\quad +\frac{(N\mathrm{+}1)(3\mathrm{+}2N\mathrm{-}2t)}
{2\, (N\mathrm{+}1/2)^2t}\cdot C^{(2)}_2 \,  \,
-\frac{(N\mathrm{+}1)(1\mathrm{-}t)}{(N\mathrm{+}1/2)^2}
\cdot \frac{dC^{(2)}_2}{dt}, 
\end{multline}
\begin{multline}
-\frac{(2N+1)^2}
{16\, (N+1)^2}  \cdot \lambda_N^2 \cdot (1-t) \cdot t^{N+1}
\, \,\, =\,\,\,\, \,
-\frac{(1-t)}{4\,(N+1)} \cdot C^{(2)}_1\\ 
\label{eqn3}
\quad \quad 
-\frac{(2\mathrm{+}2N\mathrm{-}t)(1\mathrm{-}t)}
{4\, (N\mathrm{+}1/2)^2t}\cdot C^{(2)}_2 \,\,\,
+\frac{(1\mathrm{-}t)^2}{4\,(N\mathrm{+}1/2)^2} 
\cdot \frac{dC^{(2)}_2}{dt}.
\end{multline}
From this first order coupled system we obtain 
third order uncoupled equations for the $C^{(2)}_m(N;t)$
\begin{align}
\label{c1ode}
&2\,(1-t)^2 \cdot  t^2 \cdot \frac{d^3C^{(2)}_0}{dt^3} \, \,
-6\, (N\, -(N-1)\, t)\, (1-t)\, t \cdot \frac{d^2C^{(2)}_0}{dt^2}
\nonumber\\
&\quad +2\,[N+2N^2\,+(1+4N-4N^2)\,t\, -(5N-2N^2)\,t^2] 
\cdot \frac{dC^{(2)}_0}{dt}
\nonumber\\
&\quad +(2N+1)(2Nt-2N-1) \cdot C^{(2)}_0
 \, \, =\,\, \, \,  
\frac{N\,(N+1)\,(2N+1)^2}{2} \cdot \lambda_N^2 \cdot (1-t) \cdot  t^{N}, 
\end{align}
\begin{align}
&2\,(1-t)^2\,(1+t)\cdot t^3 \cdot \frac{d^3C^{(2)}_1}{dt^3} \, \,
\, -2\,(1-t)\,[1+3N+4t+(1-3N)t^2]\cdot  t^2 \cdot \frac{d^2C^{(2)}_1}{dt^2}
\nonumber\\
&\quad+2[2+4N+2N^2 +(3+4N-2N^2) \cdot t-(3+8N+2N^2) \cdot t^2\, +2 \,N^2 \cdot t^3] \cdot  t 
\cdot \frac{dC^{(2)}_1}{dt}
\nonumber\\
\label{c2ode}
&\quad-[4+8N+4N^2 \, +(5+6N) \cdot t-(5+10N+4N^2) \cdot t^2]\cdot C^{(2)}_1
\nonumber\\
&\quad =\,\, \,  \,
\frac{(2N+1)^2\cdot [-2N^2 \, (N+1) \cdot (t+1)^2 \, +(4N+1) \cdot  t] }
{(N+1)}  \cdot \lambda_N^2 \cdot (1-t)  \cdot t^{N+1}, 
\end{align}
and
\begin{align}
&2\,(1-t)^2 \cdot  t^3 \cdot \frac{d^3C^{(2)}_2}{dt^3}\, \,  \,
-6\,(1+N-Nt) \cdot (1-t)\cdot  t^2 \cdot
\frac{d^2C^{(2)}_2}{dt^2}
\nonumber\\
&\quad+\ 2\,[7+9N+2N^2\, -(7+12N+4N^2)\cdot t\, +(1+3N+2N^2)\cdot  t^2] \cdot 
t \cdot \frac{dC^{(2)}_2}{dt}\ \ \ \ \ \ \ \ \ 
 \nonumber\\
&\quad -\ [16+24N+8N^2\, -(15+28N+12N^2)\cdot  t\, +(2+6N+4N^2)\cdot  t^2] \cdot C^{(2)}_2
\nonumber\\
 \label{c3ode}
&\quad =\,\,\, \, \frac{N^2 \,(2N+1)^4 \cdot \,(1-t)}
  {8(N+1)^2}\,  \cdot \lambda^2_N \cdot t^{N+2}. 
\end{align}

From (\ref{c1ode})--(\ref{c3ode}) it follows that  
$C^{(2)}_m(N;t)$ and
  $t^{2N+m+1} \cdot C^{(2)}_m(N;1/t)$ 
satisfy the same equation and thus, if $C^{(2)}_m(N;t)$ are
polynomials they will satisfy the palindromic property (\ref{pale}).
From (\ref{c1ode}) and (\ref{c3ode}) it follows that
the polynomials $C_0^{(2)}(N;t)$ and $C_2^{(2)}(N;t)$ satisfy 
\begin{equation}
C^{(2)}_0(N;t)\,\, \,  =\,\, \,  \, \, \, 
\frac{N}{(N+1) \cdot \beta^2_{N+1} \cdot t^2}\,  \cdot \, C^{(2)}_2(N+1;t).
\end{equation}
We therefore may restrict our considerations to $\,C^{(2)}_1(N;t)$ and 
$\,C^{(2)}_2(N;t)$.

We will obtain the polynomial solutions for the differential equations
(\ref{c1ode})--(\ref{c3ode}) by demonstrating that the homogeneous parts
  of the equations are
  homomorphic to symmetric products or symmetric powers of the second
  order operator $O_2(N)$.

\subsection{Polynomial solution for $C^{(2)}_2(N;t)$}
\label{polsol2}

Denote $\Omega^{(2)}_2(N,t)$  the order-three
linear differential operator acting
on  $\, C^{(2)}_2(N,t)$ on the left hand side of (\ref{c3ode}).
 Then it is easy to discover that the
operator $ \Omega^{(2)}_2(N,t)$  is  {\em exactly} the symmetric 
square of the second-order operator $\, O_2(N;t)$
\begin{equation}
\Omega^{(2)}_2(N,t)\,\,\,  =\,\, \, \,\, \,
\mathrm{Sym}^2\Bigl(O_2(N;t) \Bigr),
\end{equation}
which has the three linearly independent solutions 
\begin{equation}
\label{hom3}
u_1(N;t)^2,\quad \quad \quad u_1(N;t)\cdot u_2(N;t), 
\quad \quad  \quad u_2^2(N;t)
\end{equation}
where the functions $u_j(N;t)$ for $j=\,1,\,2$ are defined by
(\ref{u1def})-(\ref{hndef}).
The indicial exponents of (\ref{c3ode}) at $t=0$ are
\begin{equation}
%\label{indc21} 
2N+2,\quad\,\,  N+2,\quad\,\,  2,
\end{equation}
which are the exponents respectively of the three solutions
(\ref{hom3}).
Therefore, because the inhomogeneous term in (\ref{c3ode}) starts at
$t^{N+1}$ the coefficients $c^{(2)}_{2,n}$ in (\ref{initial}) for $0\leq n
  \leq N-1$ will be proportional to the first $N$ coefficients in the
  expansion of $u_2^2(N;t)$ about $t=0$.
 
Equation (\ref{c3ode}) is invariant under the substitution
\begin{equation}
\label{KW}
C^{(2)}_2(N;t)\,  \quad \longrightarrow\,\,\,\, \quad 
 t^{2N+3} \cdot C^{(2)}_2(N;1/t), 
\end{equation}
which maps one solution into another. Therefore if it is known that
the solution $C^{(2)}_2(N;t)$ is a polynomial the palindromic property 
\begin{equation}
c^{(2)}_{2;n}\,\,  =\, \,\,  c^{(2)}_{2;2N-1-n} 
\label{c22pal}
\end{equation}
must hold and thus $C^{(2)}_2(N;t)$ is given by
(\ref{help0}) where the normalizing constant $A^{(2)}_2$ remains to be
determined.
  
However, the invariance (\ref{KW}) is  by itself is not
sufficient to guarantee the existence of a polynomial solution with the
palindromic property (\ref{pale}). To demonstrate that there is a
polynomial solution we examine the recursion relation
which follows from (\ref{c3ode})
\begin{align}
&A^{(2)}_2 \cdot \{2n\, (2N-n)\, (N-n) \cdot c^{(2)}_{2;n}(N)
\nonumber\\
&\quad
 +(4N\,n\,-2N\,-2n^2+2n-1)\, (2n-1-2N))) \cdot c^{(2)}_{2;n-1}(N)
\nonumber\\
&\quad 
+2 \, (n-1)\, (2N-n+1)(N-n+1)) \cdot c^{(2)}_{2;n-2}(N)\}
 \, \, \,  \, \nonumber\\
&\quad = \, (\delta_{n,N}-\delta_{n,N+1}) \cdot 
\frac{N^2\, (2N+1)^4}{8\,(N+1)^2}\cdot \lambda_N^2.
\label{rr2}
\end{align}
where $c^{(2)}_{2;n}(N)=0$ for $n\leq -1$ and we may set $c^{(2)}_{2;0}=1$ by
convention . By sending $n\, \rightarrow \, 2N-n+1$ in (\ref{rr2})
 we see that $c^{(2)}_{2;n}(N)$
and $c^{(2)}_{2;2N-n-1}(N)$ 
do satisfy the same equation as required by (\ref{c22pal}). 

To prove that the solution $C^{(2)}_2(N;t)$ is indeed a polynomial we
examine the recursion relation (\ref{rr2}) for $n=N$. If there were no
inhomogeneous term then, because of the
factor $N-n$ in front of $c^{(2)}_{2;n}$, the recursion relation
(\ref{rr2}) for $n=N$ would give a constraint on $c^{(2)}_{2;N-1}$ and
$c^{(2)}_{2;N-2}$. This constraint does in fact not hold, which is the
reason that the solution $u_2^2(N;t)$ is not analytic at $t=0$ but
instead has a term $t^{N+2}\ln t$. However, when there is a nonzero
inhomogeneous term at order $t^{N+2}$ the recursion equation
(\ref{rr2}) is satisfied with a nonzero
$A^{(2)}_2$. The remaining coefficients $c^{(2)}_{2,n}$ for $N\leq
2N-1$ are determined by the palindromy constraint (\ref{c22pal}).

For $C^{(2)}_{2}(N;t)$ to be a polynomial we must have
$c^{(2)}_{2;n}(N)=0$ for $n\geq 2N$. From the recursion relation
(\ref{rr2}) we see that because of the coefficient $2N-n$ in front of
$c^{(2)}_{2;n}(N)$ the coefficient $c^{(2)}_{2;2N}(N)$ may be
freely chosen. The choice of $c^{(2)}_{2;2N}(N)\neq 0$ corresponds to
the solution of $\Omega^{(2)}_2(N;t)$ which has the indicial exponent
$N+2$ and clearly does not give a polynomial solution. However 
 by setting $n=2N+1$ in (\ref{rr2})  
we obtain
\begin{equation}
\label{deux}
2 \, (N+1) \, (2\, N+1) \cdot c^{(2)}_{2; 2N\, +1}(N)
\,\, \, -\,\,\,(2\, N+1)^2 \cdot   c^{(2)}_{2;2N}(N)\,\, = \,\, \, 0.
\end{equation}
and if we
choose $c^{(2)}_{2;2N}(N)=0$ we obtain
$c^{(2)}_{2;2N+1}(N)=0$ also. Therefore because (\ref{rr2}) is a three
term relation, it follows that $c^{(2)}_{2;n}(N)=0$ for $n\geq 2N$
as required for a polynomial solution.

It remains to explicitly evaluate the normalization constant $A^{(2)}_2$
which satisfies (\ref{rr2}) with $n=N$. 
A more efficient derivation is obtained
if we return to the original inhomogeneous equation (\ref{c3ode}).
Then we  note that if we include the term with
$n=0$ in the second terms on the right-hand side of (\ref{u2def}) 
in the computation of the
term of order $t^{N+2}$ in the left hand side of (\ref{c3ode}) we must
get zero because $u_2^2$ is a solution of the homogeneous part of
(\ref{c3ode}). Therefore when we use the extra term in $u_2^2$ of
\begin{equation}
-2 \,\,t^{N+2} \cdot N \cdot \lambda^2_N \cdot \ln t, 
\end{equation}
in the lhs of (\ref{c3ode}), and keep the terms which do not involve
$\ln t$, we find
\begin{equation}
\label{odesum}
2\, (N^2-1)\cdot c^{(2)}_{2;N-2}(N)\,\,\,  
-(2N^2-1)\cdot c^{(2)}_{2;N-1}(N) \, \, =\,\, \,
  -4\, N^3 \cdot \lambda_N^2. 
\end{equation}
Thus, using  (\ref{odesum}) we evaluate
(\ref{rr2}) with $n=N$ as
\begin{equation}
-4 \,A^{(2)}_2 N^3\,\lambda_N^2 \, \,\,\, 
= \, \, \,\,\, \, \frac{N^2\, (2N+1)^4 }{8\,(N+1)^2} \cdot \lambda_N^2, 
\end{equation}
and thus
\begin{equation}
A^{(2)}_2 \, \,\,  = \, \,\, \,\,  -\frac{N}{2} \cdot \beta_N^2.
\end{equation}

\subsection{Polynomial solution for $C^{(2)}_1(N;t)$}
\label{polsol1}

The computation of $C^{(2)}_1(N;t)$ has features which are
characteristic of $C^{(n)}_m(N;t)$  which 
are not seen in  $C^{(2)}_2(N;t)$. Similarly to what has
been done in the previous subsection we introduce 
$\Omega^{(2)}_1(N;t)$, the order-three linear differential operator
acting on $\,  C^{(2)}_1(N;t)$ in the lhs of (\ref{c2ode}).

The indicial exponents at $t=0$ of the operator 
$\Omega^{(2)}_1(N;t)$  are 
\begin{equation}
\label{indc21} 
1,\quad\,\,  N+1,\quad\,\,  2N+2 
\end{equation}
This  order-three operator $\Omega^{(2)}_1(N;t)$
is found to be related to the {\em symmetric product}
 of 
$O_2(N)$ and $O_2(N+1)$ by the direct sum decomposition
\begin{equation}
\label{decomp}
\mathrm{Sym}\Bigl(O_2(N),\, O_2(N+1)\Bigr) \cdot t
 \,\, \,\,  = \, \, \,\,\,\, \, 
\Omega^{(2)}_1 \oplus \Bigl( D_t\, \,  \, \, -\frac{N+1}{t}\Bigr). 
\end{equation}
The three linearly independent solutions of 
$\Omega^{(2)}_1(N;t)$ are to be found in the set of four functions
\begin{align}
\label{set1}
&t^{-1} \cdot u_1(N;t)\cdot u_1(N+1;t),  
&t^{-1}\cdot u_2(N,t)\cdot u_1(N+1;t), & \quad
\nonumber\\
&t^{-1} \cdot u_1(N;t)\cdot u_2(N+1;t), 
&t^{-1} \cdot u_2(N;t) \cdot u_2(N+1;t), & \quad
\end{align}
where from the definitions of $u_1(N;t)$ in (\ref{u1def}) and $u_2(N;t)$ in 
(\ref{u2def}) the behaviors of these four solutions as $t\, \rightarrow\, 0$ 
are $t^{2N+2}, \,t^{N+2}, \,t^{N+1}, \,t$ respectively.

Following the argument given above for $C^{(2)}_2(N;t)$ 
we conclude that because the inhomogeneous term in (\ref{c1ode})
is of order $t^{N+1}$ that the terms up through order $t^N $ must be
proportional to the solution of the homogeneous equation
\begin{equation}
t^{-1} \cdot u_2(N;t) \cdot u_2(N+1;t), 
\end{equation} which begins at order $t$. This observation determines
the form (\ref{initial}) and the coefficients (\ref{dn1}) 
$c^{(2)}_{1;n}(N)$ for
$0\leq n\leq N-1$. The normalizing constant $A^{(2)}_{1}$ and the
remaining coefficient $c^{(2)}_{1;N}(N)$ (\ref{c21n}) are then
obtained from the inhomogeneous equation (\ref{c1ode}). Finally, to prove that
$C^{(2)}_{1}(N;t)$ is actually a palindromic polynomial the recursion 
relation for the coefficients $c^{(2)}_{1;n}(N)$ must be used. Details
of these computations are given in \ref{polsol1ap}.

\subsection{The constant $K^{(2)}_0$}
\label{constK02}

Finally, we need to evaluate the constant of integration $K^{(2)}_0$ in
(\ref{feform}). This is easily done by noting that from the original
integral expression (\ref{dffm}) that $f^{(2)}_{N,N}(0)=\, 0$ for all $N$.
From (\ref{initial})--\,(\ref{final}) we see that
\begin{equation}
C^{(2)}_0(N;0) \,\,  = \, \,  -\frac{N}{2},
\quad \quad \quad 
C^{(2)}_1(N;0)
 \, \,  =  \,  \, \, C^{(2)}_2(N;0) 
\, \,  = \,  \, \,  0, 
\end{equation}
and using this in (\ref{feform}) we obtain
$\, K^{(2)}_0 \, \,  = \,  \, \,  N/2\, $ 
as desired.

\section{The derivation of the results for  $f^{(3)}_{N,N}(t)$}
\label{ffactor}

The form factor $f^{(3)}_{N,N}(t)$ is defined by the integral
(\ref{dffp}) with $2n+1 \, = \, \,  3$,
 and if we are to follow the method of
evaluation developed for $f^{(2)}_{N,N}(t)$, we need to demonstrate
analytically that there is an  operator
 which, when acting on the integral,
will split it into three factors. Unfortunately we have not analytically
obtained such a result. 

However, we are able to proceed by using the methods of differential
algebra and
from~\cite{mccoy1}
it is known computationally for integer $N$ that $f^{(3)}_{N,N}$ 
is annihilated by the operator $L_4(N)\cdot L_2(N)$ where
\begin{equation}
L_2(N)  \,  \, = \, \, \,   \,  D_t^2 \, \, \, 
+\frac{2t-1}{(t-1)t}\cdot D_t \,\, 
 -\frac{1}{4t} \,\,  +\frac{1}{4(t-1)} \, -\frac{N^2}{4t^2}, 
\end{equation}
and $L_2(N)$ annihilates $f^{(1)}_{N,N}(t)$, and where,
\begin{align}
&L_4(N)\, \, = \,\,\,\,  D_t^4\,\, \, 
+10\frac{(2t-1)}{(t-1)\, t} \cdot D_t^3 \,\, \, 
+\frac{(241t^2-241t+46)}{2\, (t-1)^2\, t^2} \cdot D_t^2
\nonumber\\
&\quad +\frac{(2t-1)(122t^2-122t+9)}{(t-1)^3\,t^3} \cdot D_t\,\, 
+\frac{81}{16}\frac{(5t-1)(5t-4)}{t^3\,(t-1)^3}\, 
-\frac{5}{2}\frac{N^2}{t^2} \cdot D_t^2
\nonumber\\
&\quad +\frac{(23-32t)N^2}{2\, (t-1)\, t^3} \cdot D_t\, \, 
+\frac{9}{8}\, \frac{(8-17t)\, N^2}{(t-1)\, t^4}\, \, 
+\frac{9}{16}\, \frac{N^4}{t^4}.
\end{align}
Furthermore the operator $L_4(N)$ is homomorphic to the symmetric 
cube of $L_2(N)$ by the following relation,
\begin{equation}
L_4(N)\cdot Q(N)\,\,  \,\,=\,\,\,\,\,  \,  R(N)\cdot \mathrm{Sym}^3(L_2(N)), 
\end{equation}
where,
\begin{multline}
Q(N)\,\, =\,\,\,\,\,  
  (t-1)\cdot  t \cdot D_t^3\, \, 
+\frac{7}{2}(2t-1)\cdot D_t^2\,\,
  +\frac{(41t^2-41t+6)}{4\, (t-1)\, t}\cdot D_t
\\
 +\frac{9}{8}\frac{(2t-1)}{(t-1)\, t}\,\,
 -\frac{9}{4}\frac{(t-1)N^2}{t} \cdot D_t\, \,
-\frac{9}{8}\frac{(2t-1)}{t^2}\,N^2, 
\end{multline}
and
\begin{align}
R(N) \,\,  &=\,\, \, \,  (t-1)\cdot  t \cdot D_t^3\,
 +\frac{23}{2}\, (2t-1) \cdot D_t^2\, \, 
+\frac{21}{4}\, \frac{6-29t+29t^2}{(t-1)\, t} \cdot D_t
\nonumber\\
&\qquad+\frac{9}{8}\frac{(2t-1)\, (125t^2-125t+16)}{(t-1)^2\, t^2}\,\,  
-\frac{9\, N^2}{4} \cdot \left(\frac{(t-1)}{t}  \cdot D_t \, 
+\,\frac{(10t-9)}{2 \, t^2} \right). 
\end{align}

We therefore conclude that since $f^{(3)}_{N,N}(t)$ is regular at
$t=\, 0$ and the solution of $L_2(N)$ which is regular at $t=\, 0$
is $F_N$, that
\begin{equation} 
\label{ANrelation}
Q(N) \cdot  B_0  \cdot t^{3N/2} \cdot F_N^3
 \, \, \, =  \, \, \,\,  \,L_2(N) \cdot f^{(3)}_{N,N}, 
\end{equation}
where $B_0$ is a normalizing constant which is determined from the
behavior at $t\, =\,\,  0$. From the integral (\ref{dffp}) we find   
\begin{equation}
f^{(3)}_{N,N} \, \, =\, \,  \, \,\, 
 \frac{N+2}{4\, (N+1/2)}\, \left(\frac{(1/2)_{N+1}}{(N+2)!^3}\right)^3 
\cdot t^{3N/2+2} \,\, \, \,  +\, O(t^{3N/2+3}),
\end{equation}
and from the expansion of $F_N$ we have
\begin{equation}
Q(N)\cdot t^{3N/2}\cdot F^3_N
\, \,\, =\,\,\,\, \,  
  \frac{3(2N+1)^3}{8\, (N+1)^2\, (N+2)}\cdot t^{3N/2} 
\,\, \, \,  +\, O(t^{3N/2+1}), 
\end{equation} 
and thus
\begin{equation}
\label{b0def}
B_0\,\, \,=\,\,\,\, \, \frac{1}{3} \cdot \lambda_N^3.
\end{equation}

Operating $Q(N)$ on $ \, t^{3N/2}F_N^3$, one can write the result in the 
basis $F_N$ and $\bar{F}_N$. Similarly, one can operate on the form
${f}^{(3)}_{N,N}$ in (\ref{foform}) with $L_2(N)$ and write the result in the same 
basis $F_N$ and $\bar{F}_N$.
Then, matching powers of the hypergeometric functions on both 
sides of the relation (\ref{ANrelation}) will 
yield four coupled inhomogeneous 
ODEs to be solved. 
The four coupled ODEs are given in  \ref{appcoupled}.

For $C^{(3)}_{m}(N;t)$ with $m=\, 0,1,2$, the reduction of the  
four coupled second order equations leads to inhomogeneous 8-th order 
uncoupled ODEs  for each $C^{(3)}_m(N;t)$ separately, of the form
\begin{equation}
\sum_{j=0}^8\, P_{m,j}\,(t)\cdot  t^j \cdot  \frac{d^j}{dt^j}C^{(3)}_m(N;t)
\,\, \,\,  =\,\,\, \,\,   I_m(t),
\end{equation}
where
\begin{align}
&I_0\,\,  =\,\,\, \,   t^{N+1} \cdot \sum_{j=0}^{14}\, I_0(j) \cdot t^j, 
\qquad \quad 
I_1\,\,  =\,\,\, \,   t^{N+1} \cdot \sum_{j=0}^{17}\, I_1(j) \cdot t^j,
\nonumber\\
&I_2\,\,  =\,\,\,\,    t^{N+2} \cdot \sum_{j=0}^{14}\, I_2(j) \cdot t^j,
\end{align}
where the $I_m(t)$ are antipalinromic and $P_{m,n}(t)$ are polynomials. In 
particular
\begin{equation}
P_{m,8}(t)\,\,\,   =\,\, \, \, \,  (1-t)^9 \cdot P_m(t),
\end{equation}
where $P_0(t)$ and $P_2(t)$ are order six and $P_1(t)$ is order eight.

However, for
$C^{(3)}_3(N;t)$ a step-by-step elimination process in the coupled system terminates
 in a fifth order equation instead. We derive and present this 5th order equation in
  \ref{appc3},  but the eighth order equations given by Maple
 are too long to present.

\subsection{Polynomial solution for $C^{(3)}_3(N;t)$}
\label{polsolC3}

The homogeneous operator on the LHS of the ODE (\ref{c33ode}) for 
$C^{(3)}_3(N,t)$ is found on Maple to be isomorphic 
to  ${\rm Sym}^4(O_2(N)) \cdot t^{(N+1)}$,
the symmetric fourth power of $O_2(N)$ multiplied by $t^{(N+1)}$. Therefore all five
 solutions of the homogeneous equation are 
given as $t^{-(N+1)}$ times products of the solutions $u_1(N;t)$ and
$u_2(N;t)$. The fifth order ODE has at $t\,=\,\, 0$ the indicial exponents
\begin{equation}
\label{indicialc3}
-N+3,\quad \, \, 3, \quad \, \, N+3, \quad \, \,2N+3, \quad \, \, 3N+3.
\end{equation}
Therefore because the polynomial solution must by definition be
regular at $t=\, 0$ the first $N+1$  terms (from $t^3$ through $t^{N+3}$)
in the solution
\begin{equation}
t^{-(N+1)} \cdot u_2^3(N)\cdot u_1(N),
\label{c33sol}
\end{equation} 
which vanishes as $t^3$, will 
solve the inhomogeneous equation
(\ref{c33ode}), so that 
\begin{equation}
\label{c33polysec4}
C^{(3)}_3(N;t)\,\, =\, \, \, \,\, \, \,A^{(3)}_3 \cdot 
t^3 \cdot \sum_{n=0}^{2N-2}\, c^{(3)}_{3;n}\cdot t^n, 
\end{equation}
where for $0 \,\leq \, n \, \leq \, N-1$
\begin{equation}
\label{c33nsec4}
c^{(3)}_{3;n}\, \,=\,\,\,  \,\, 
\sum_{m=0}^n\sum_{l=0}^m\sum_{k=0}^l\, \, a_k(N)
\cdot a_{l-k}(N) \cdot a_{m-l}(N) \cdot b_{n-m}(N).
\end{equation}
The lowest order inhomogeneous term is $t^{N+3}$ which is the next
indicial exponent in (\ref{indicialc3}) and therefore the normalizing
constant $A^{(3)}_3$  is found from the first logarithmic term in the
solution of the homogeneous equation by exactly the same argument used for 
$C^{(2)}_2(N;t)$. Thus we find 
\begin{equation}
A^{(3)}_3\,\,  =\, \, \, \, \,\frac{2}{3} \cdot \beta_N^3 \cdot \lambda_N.
\end{equation}

The remaining demonstration that $C^{(3)}_3(N;t)$ is a palindromic
polynomial follows from the recursion relation for the 
coefficients, as was done for
$C^{(2)}_2(N;t)$, with the exception that because the inhomogeneous
term in (\ref{c33ode}) is proportional to $t^{N+1}(t^2-1)$ instead of
$t^{N+1}(t-1)$, there is an identity which must be verified.
Details are given in \ref{appc3}.

\subsection{Polynomial solutions for $C^{(3)}_2(N;t)$ and  $C^{(3)}_0(N;t)$. }
\label{polsolC3C3}

A new feature appears in the computation of  
$C^{(3)}_2(N;t)$ and  $C^{(3)}_0(N;t)$.

The indicial exponents at $t=0$ of
the 8-th order operator  $\Omega^{(3)}_2(N;t)$ 
\begin{align}
\label{indicial32}
-N+2,\,\,\, 2, \,\,\, 3, \,\,\, N+2, \,\,\, N+3, \,\,\,
 2N+2, \,\,\, 2N+3, \,\, \,3N+3,
\end{align}
and for  $\Omega^{(3)}_0(N;t)$ are
\begin{equation}
\label{indicial30}
-N, \quad 0, \quad 1, \quad N+1, \quad N+2, \quad 2N+2, 
\quad 2N+3, \quad 3N+3,
\end{equation}
and from these exponents it might be expected that the solution of 
$\Omega^{(3)}_2(N;t) ~~(\Omega^{(3)}_0(N;t))$ which is of order
$t^2~~(t^0)$ could have a logarithmic term $t^3\ln t~~(t\ln t)$ which
would preclude the existence of a polynomial solution of the
corresponding inhomogeneous equation. However, this does, in fact, not
happen because there is a decomposition of the 8-th order
operators into a direct sum of the third order operators
$\Omega^{(2)}_2(N;t)~~(\Omega^{(2)}_0(N;t))$ with exponents 
$2,~N+2,~2N+2~(0,N+1,2N+2)$ and new
fifth order operators $M^{(3)}_m(N;t)$
\begin{align}
\Omega^{(3)}_m(N;t)\,\,= \,\,\, M^{(3)}_m(N;t)\, \oplus \, \Omega^{(2)}_m(N;t) 
\label{dirsum3}
\end{align}
with exponents $-N+2,~2,~N+2,~2N+2,~3N+2$ for $M^{(3)}_2(N;t)$ and
$-N,~0,~N+1,~2N+2,~3N+3)$ for $M^{(3)}_0(N;t)$.
Furthermore $M^{(3)}_2(N;t)$ is homomorphic
to the symmetric fourth power of $O_2(N)$ and 
$M^{(3)}_0(N;t)$ is homomorphic
to the symmetric fourth power of $O_2(N+1)$
 (see \ref{diffop} for details).
The inhomogeneous equation is solved in terms of a linear
combination of the solutions of the third order and fifth order
homogeneous equations.

However, a simpler form of the answer results if we notice the
isomorphisms
\begin{align}
\Omega^{(3)}_2(N;t)
\,\,  \, &=\, \, \,  \, \,\, 
\mathrm{Sym}\left(O_2(N),O_2(N),O_2(N), O_2(N+1)\right)\cdot t^{N+2}, 
\\
\Omega^{(3)}_0(N;t)
\,\,  \, &=\, \, \,  \, \, \, 
\mathrm{Sym}\left(O_2(N), O_2(N+1),O_2(N+1),O_2(N+1)\right) \cdot t^{N+4}, 
\end{align}

The desired solutions for $\Omega^{(3)}_2(N;t)$ are constructed 
from the two solutions which have the exponents  2 and 3, 
\begin{equation}
t^{-N-2} \cdot u_2^2(N)\cdot u_1(N)\cdot u_2(N+1),
 \qquad t^{-N-2} \cdot u^3_2(N) \cdot u_1(N+1),
\end{equation}
which by use of the Wronskian condition (\ref{u1u2})
may be rewritten as  a linear combination of two solutions each with
the exponent of 2 as 
\begin{equation}
A^{(3)}_2\cdot t^{-N-2} \cdot u_2^2(N)\cdot u_1(N)\cdot u_2(N+1)\, \, \, \,
 +B^{(3)}_2 \cdot u_2^2(N),
\end{equation}
and similarly for  $C^{(3)}_0(N;t)$, 
we choose as the
solution of the homogeneous equation the two solutions with exponent $0$
\begin{equation}
A^{(3)}_0 \cdot t^{-N-4} \cdot u_2^3(N+1) \cdot u_1(N) \,
\, \, +B^{(3)}_0 t^{-2}\cdot u_2^2(N+1).  
\end{equation}
This procedure determines the constants $c^{(3)}_{2;n}$ for $0\leq n \leq N-1$ and 
and $c^{(3)}_{0;n}$ for $0\leq n \leq N$, with palindromy
 determining the remaining $c^{(3)}_{2;n}$ for $N\leq n \leq 2N-1$ (\ref{c32n})
 and and $c^{(3)}_{0;n}$ for $N+1\leq n \leq 2N+1$ (\ref{c30n}).

The constants  
$A^{(3)}_{2}$ and $ B^{(3)}_{2}$ in (\ref{c02form})
are found by using (\ref{c02form}) with (\ref{c32n}) in the
inhomogeneous equation for $C^{(3)}_2(N;t)$ and matching the first two
terms in the inhomogeneous terms of orders $t^{N+2}$ and $t^{N+3}$
(which are the same orders as the corresponding indicial exponents 
(\ref{indicial32})). This generalizes the determination
 of $A^{(2)}_m$ for $C^{(2)}_m$ above. 
Similarly the constants $A^{(3)}_{0}$ and
$B^{(3)}_{0}$ are found using  (\ref{c02form}) with (\ref{c30n}) 
in the inhomogeneous equation for $C^{(3)}_0$
and matching to the inhomogeneous terms $t^{N+1}$ and $t^{N+2}$.
Thus we obtain the results (\ref{c02form})-(\ref{a3n})  
summarized in section \ref{explicit}.
 
\subsection{Polynomial solution for $C^{(3)}_1(N;t)$}
\label{polsolC31}

The computation of $C_1^{(3)}(N;t)$ has further new features.

The 8-th order homogeneous operator $\Omega^{(3)}_1(N;t)$ of the inhomogeneous
equation for $C^{(3)}_1(N;t)$ 
has the eight indicial exponents at $t=0$
\begin{equation}
-N+1,\, \, \,  1, \, \, \, 2, \, \, \, N+1,
\, \, \, N+2, \, \, \, 2N+2, \, \, \, 2N+3, \, \, \, 3N+3,
\end{equation}
and, as in the case of $\Omega^{(3)}_0(N;t)$ and $\Omega^{(3)}_2(N;t)$
has a decomposition into a direct sum of $\Omega^{(2)}_1(N;t)$ and a fifth
order operator. However, simpler results are obtained by observing that
$\Omega^{(3)}_1(N;t)$ is homomorphic to the symmetric product 
\begin{eqnarray}
&&\mathrm{Sym}\Bigl(O_2(N),\, O_2(N),\, O_2(N+1),O_2(N+1)\Bigr) \cdot
t^{N+3} \nonumber \\
&&\qquad \qquad = \, \, \,
 \Omega^{(3)}_1(N;t) \oplus \Bigl(D_t-\frac{(N+1)}{t}\Bigr),
\end{eqnarray}
 which
satisfies a 9-th order ODE with indicial exponents at $t=\,0$
 \begin{equation}
-N+1, \, \, \, 1, \, \, \, 2, \, \, \, N+1, \, \, \, N+2,
 \, \,\,  N+3, \, \, \, 2N+2, \, \,\,  2N+3, \, \, \,3N+3.
\end{equation}
The solutions with exponents of 1 and 2 are respectively
\begin{equation}
t^{-N-3} \cdot u_2(N) \cdot u_1(N) \cdot u_2^2(N+1), \quad\,\,
 t^{-N-3} \cdot u_2^2(N) \cdot u_2(N+1) \cdot u_1(N+1),
\end{equation} 
and again, recalling the Wronskian relation (\ref{u1u2}),
we may construct the polynomial $C^{(3)}_1(N;t)$, similar to the
construction of $C^{(3)}_{2}(N;t)$, from the linear
combination 
\begin{equation}
A^{(3)}_{1} \cdot t^{-N-3} \cdot u_2(N) \cdot u_1(N) \cdot u_2^2(N+1)\, \, 
 +\,\,B^{(3)}_1 \cdot t^{-1} \cdot u_2(N) \cdot u_2(N+1), 
\label{b31}
\end{equation}
which determines the coefficients $c^{(3)}_{1;n}$ for 
$0\leq n \leq N-1$, with palindromy determining the remaining 
$c^{(3)}_{1;n}$ for $N+1\leq n \leq 2N$ (\ref{c31n}).
The coefficients $A^{(3)}_1$ and $B^{(3)}_1$ (\ref{b31})
are determined in a manner similar to the determination of 
$A^{(3)}_2$ and $B^{(3)}_2$,
 by matching to the terms of order $t^{N+1}$ and $t^{N+2}$.

Finally the term $c^{(3)}_{1;\, N}$ 
is computed by using the previously  determined results for 
$C^{(3)}_2(N;t)$ and $C^{(3)}_3(N;t)$ in the 
coupled differential equation (\ref{coupled4}), giving the result 
(\ref{c31middle}).
 
\subsection{Determination of $K^{(3)}_0$}

It remains to determine the constant $K^{(3)}_0$ (\ref{k30}), 
which is easily done by setting $t=0$ in (\ref{foform}) 
to obtain
\begin{equation}
0\,=\,\,\,K^{(3)}_0 \cdot \lambda_N \, +A^{(3)}_0 \,
 +\frac{N-1}{N} \cdot \lambda_N \cdot  A^{(2)}_0,
\end{equation} 
and using (\ref{final}) and (\ref{a3n}). 

\section{The Wronskian cancellation for $f^{(2)}_{N,N}(t)$ 
and $f^{(3)}_{N,N}(t)$ }
\label{wronsk}

%We introduce,  here, a wronskian cancellation method {\em in order to get 
%exact results for arbitrary values of $\, N$}.

The polynomials $C^{(n)}_m(N;t)$ are of order $t^m$ as $t\rightarrow
0$. However, from (\ref{f2t0}) and (\ref{f3t0}) we see that
$f^{(2)}_{N,N}(t)$ vanishes as $t^{N+1}$ and $f^{(3)}_{N,N}(t)$ 
vanishes as $t^{N+2}$. Therefore for $t\rightarrow 0$, a great deal of
cancellation must occur in (\ref{feform}) and (\ref{foform}). This
cancellation is an important feature of the structure of the results
of sec. 2.2 and 2.3.

{\em To prove the cancellations} we note that the $n$-th power of the
Wronskian relation (\ref{u1u2}) is
\begin{equation}
t^{-n(N+2)} \cdot  \sum_{j=0}^n (-1)^j \cdot   \binom{n}{j}  \cdot  \beta_N^j \cdot 
 \left[u_2(N+1)u_1(N)\right]^{n-j} \cdot \left[u_2(N)u_1(N+1)\right]^j
 \,\,  =\, \, \,\,   1,
\label{wron2}
\end{equation}
or alternatively,
\begin{equation}
\sum_{j=0}^n (-1)^j \cdot  \binom{n}{j}\cdot   \beta_N^j   \cdot 
\left[\frac{u_2(N+1)}{t}\right]^{n-j} \cdot  u_2(N)^j \cdot F_N^{n-j} \, F_{N+1}^j 
\,\,  = \, \,\, \,  1.
\label{wron3}
\end{equation}
Thus, by defining $\stackrel{N}{=}$ to mean equality up though and
including terms of order $t^N$ we see immediately from the form
(\ref{feform}) with (\ref{cdef}) for $K^{(2)}_0$ and (\ref{help0}),
(\ref{dn1}) and (\ref{c2n0}) for the $c^{(2)}_{m,n}$ with $0\leq n\leq N$
that
the terms though order $t^N$ in $f^{(2)}_{N.N}(t)$ are
\begin{equation}
f^{(2)}_{N,N}  \,  \stackrel{N}{=}\, \, \,\,  
  \frac{N}{2} \cdot \left\{1-\sum_{j=0}^2 (-1)^j
\binom{2}{j} \cdot \beta_N^j \cdot   
\left[\frac{u_2(N+1)}{t}\right]^{2-j} \cdot u_2(N)^j~
F_N^{2-j}F_{N+1}^j \right\},
\label{f2v}
\end{equation}
which vanishes by use of (\ref{wron3}). This derivation has made no
use of $c^{(2)}_{1;N}$. This term contributes only to order $t^{N+1}$
and may be determined from the normalization amplitude
(\ref{f2t0}). This provides an alternative to the derivation of
(\ref{c21n}) of Appendix B. 

To prove the cancellation for $f^{(3)}_{N,N}(t)$
we note that because of the term $C^{(2)}_m(N;t)$ in  $C^{(3)}_m(N;t)$
for $m=0,1,2$ in (\ref{c02form}) we may use the expression
(\ref{feform}) and (\ref{cdef})-(\ref{c2n0}) for $f^{(2)}_{N,N}(t)$ in the form
\begin{equation}
\sum_{m=0}^2C^{(2)}_m(N;t)\cdot F_N^{2-m} \, F_{N+1}^m\,\,  = \,\, \,
 f^{(2)}_{N,N}(t)\,\, \, -\frac{N}{2}.
\label{aux1}
\end{equation}
Thus from (\ref{foform}), (\ref{f1}) and (\ref{aux1}) we obtain an
alternative form for $f^{(3)}_{N,N}(t)$ of
\begin{equation}
f^{(3)}_{N,N}(t)\,=\,\,\,  \left\{\frac{2}{3} \,   \,  \, 
+\frac{N-1}{N}f^{(2)}_{N,N}(t)\right\} \cdot f^{(1)}_{N,N}(t)\,\,\,\, 
+t^{N/2} \cdot \sum_{m=0}^3{\bar C}_m^{(3)}(N;t)\cdot F^{3-m}_N \,  F^m_{N+1},
\label{f3alt}
\end{equation}
where
\begin{equation}
{\bar C}_m^{3}(N;t)
\, =\,\, \, \,  (-1)^{n+1} \cdot \frac{2}{3} \cdot {n\choose 3}
 \cdot \beta_N^n \cdot \lambda_N \cdot 
\sum_{n=0}^{2N+1-m}c^{(3)}_{m;n}t^n.
\end{equation}
We have already demonstrated by use of (\ref{f2v}) that
$f_{N,N}^{(2)}(t)$ vanishes though order $t^N$. Therefore using the
expressions (\ref{c33n})-(\ref{c31n}) for $c^{(3)}_{m;n}$ which are 
all valid through (at least)  order $t^N$ and the definition (\ref{f1}) 
of $f^{(1)}_{N,N}(t)$ we find 
\begin{equation}
\frac{f^{(3)}_{N,N}(t)}{t^{N/2}} \, \stackrel{N}{=}\,\,  \frac{2}{3}\lambda_NF_N \cdot \left\{1 \, \, 
-\sum_{j=0}^3 (-1)^j \cdot \binom{3}{j} \cdot \beta_N^j  \cdot 
\left[\frac{u_2(N+1)}{t}\right]^{3-j} \cdot u_2(N)^j \cdot F_N^{3-j}\, F_{N+1}^j\right\}, 
\label{f3v}
\end{equation}
which vanishes by use of the Wronskian relation (\ref{wron3}) with $n=3$. 

We have thus demonstrated that $f^{(3)}_{N,N}(t)/t^{N/2}$ vanishes to order
$t^N$ as $t\rightarrow 0$. However we see, from the original integral
(\ref{dffp}), that in fact $f^{(3)}_{N,N}(t)/t^{N/2}$ is of order
$t^{N+2}$. Therefore the coefficient of $t^{N+1}$ must also
vanish. This is not proven by (\ref{f3v}). However the coefficient
$c^{(3)}_{1,N}$ has not been used in the derivation of (\ref{f3v}) and
the choice of  $c^{(3)}_{1,N}$ to make the coefficient of $t^{N+1}$
vanish provides an alternative derivation of (\ref{c31middle}). 
 
\section{Factorization for $f^{(n)}_{N,N}$ with $n\geq4$} 
\label{formfac}

In principle the methods  of differential algebra of the previous 
sections can be extended to
form factors  $f^{(n)}_{N,N}(t)$ with $n\geq 4$. However, the complexity
of the calculations rapidly increases. 

For $f^{(2n)}_{N,N}(t)$ there are $2n+1$ polynomials $C^{(2n)}_m(N;t)$
and since from~\cite{mccoy1} we find that for $N\, \geq \, 1$
\begin{equation}
L_{2n+1}\,  \cdots \,  L_3\cdot L_1 \cdot f^{(2n)}_{N,N}(t)
\,\,\,\, \, =\,\,\, \,\,\,0,
\end{equation}
where $L_k$ is a linear differential operator of order $k$,
the polynomials $C^{(2n)}_m(N;t)$ will satisfy a system of $2n+1$
coupled differential equations where the maximum derivative order is
$n^2$. These equations can be decoupled into $2n+1$ Fuchsian ODEs which
generically have order $n^2\,(2n+1)$.

 Similarly for $f^{(2n+1)}_{N,N}(t)$ we found in~\cite{mccoy1} that
\begin{equation}
L_{2n+2} \, \cdots  \, L_4  \cdot   L_2 \cdot f^{(2n+1)}_{N,N}(t)
\,\,\,\,=\,\,\,\,\, 0,
\end{equation} 
and thus
the $2n+2$ polynomials
$C^{(2n+1)}_m(N;t)$ satisfy inhomogeneous coupled equations of 
maximum differential order $n(n+1)$ which 
for $N\geq 1$ are generically decoupled into 
Fuchsian equations of order $2n(n+1)^2$.

We have obtained for $f^{(4)}_{N,N}(t)$ the 20-th order ODEs for
$C_m^{(4)}(N;t)$ in the cases $N \,=\,\,1,\cdots, 10$ and will 
illustrate the new features which arise by considering the case $m=4.$
 
We find by use of Maple that (at least for low values of $N$)
the operator $\Omega^{(4)}_4(N;t)$ has a
direct sum decomposition
\begin{equation}
\Omega^{(4)}_4(N;t)\,\,\, \, =\,\,\,\,\, \, M^{(4)}_7(N)\oplus M^{(4)}_{5;1}(N)
\oplus M^{(4)}_{5,2}(N) \oplus M^{(4)}_3(N),
\label{sirsim4}
\end{equation}
where $M^{(4)}_{k;n}(N)$ is order $k$ and is 
homomorphic to the symmetric $k-1$ power of $O_2(N)$ 
\begin{align}
M^{(4)}_7(N)\cdot J^{(4)}_2(N;t)\,\,&=\,\,\, G^{(4)}_{2}(N;t)
\cdot{\rm Sym}^6(O_2(N)),\label{m7}\\
M^{(4)}_{5;1}(N)\cdot J^{(4)}_1(N;t)\,\, 
&=\,\,\, G^{(4)}_1(N;t)\cdot {\rm Sym}^4(O_2(N)),
\label{m51}\\
M^{(4)}_{5;2}(N)\,\, &=\,\,\, {\rm Sym}^4(O_2(N)),
\label{m52}\\ 
M^{(4)}_3(N)\cdot J_0(N;t)\,\,
&=\,\,\, G^{(4)}_0(N;t) \cdot{\rm Sym}^2(O_2(N)), 
\label{m3} 
\end{align}
where the intertwinners $\,J^{(4)}_m(2;t)$ and  $\,G^{(4)}_m(2;t)$
are linear differential operator of order $\, m$. The
 intertwinners $\,J^{(4)}_m(2;t)$ in (\ref{m7})-(\ref{m3}), are
explicitly  given in \ref{more}. 
Further examples of intertwinners are given in \ref{more}.
These differential algebra exact results (in particular 
(\ref{m7})-(\ref{m3})) are the illustration of the 
{\em canonical link between the palindromic polynomials 
 and the theory of elliptic curves}.

Direct sum decompositions\footnote[1]{Note that in direct sum decomposition 
like (\ref{sirsim4}), some ambiguity may occur with terms like 
$\, M^{(4)}_{5;1}(N) \oplus M^{(4)}_{5,2}(N)$ where $\,M^{(4)}_{5,1}(N)$ 
and   $\,M^{(4)}_{5,2}(N)$ are both homomorphic to a same operator
 (here $\,{\rm Sym}^4(O_2(N)) $).} have been obtained for 
$\Omega^{(2)}_m(N;t),~\Omega^{(3)}_m(N;t)$ and $\Omega^{(4)}_m(N;t)$
and {\em we conjecture that this occurs generically
for all} $\Omega^{(n)}_m(N;t)$. Taking into account the
homomorphism of  $O_2(N;t)$ and $O_2(N+1;t)$, and recalling,
for instance,
 subsections \ref{polsolC3C3}) and \ref{polsolC31}, it may be easier
to write direct sum decomposition formulae in terms of  sum of 
symmetric products of $O_2(N;t)$ and $O_2(N+1;t)$. In order to extend
 these results, beyond these few
special cases of $\Omega^{(4)}_m(N;t)$, a deeper and systematic study of
the homomorphisms is still required.

From an analytical viewpoint, a complication which needs to be understood 
is how to use the solutions
of the homogeneous operators $\Omega^{(n)}_m(N;t)$ to obtain the
polynomial solution of the inhomogeneous equations. The first
difficulty here is that for $C^{(4)}_m(N;t)$ the inhomogeneous terms 
are large polynomials, of order 100 and higher. Moreover, the orders
of palindromy point of the
 $C^{(4)}_4(N;t)$ with $N=1,\cdots,10$ are 
 all larger than the order $t^{N+4}$ where the solutions of
the homogeneous operators $M^{(4)}_{k;n}(N;t)$ have their first logarithmic
singularity. Consequently linear combinations of solutions must be
made which cancel these logarithmic singularities at $t^{N+4}$ to give sets of
solutions to $\Omega^{(4)}_4(N;t)$ which are analytic up to the order
of the first inhomogeneous terms. Thus the determination of the
correct linear combination of solutions of the operators
$M^{(4)}_{k;n}(N;t)$ is significantly more complex than was the case 
for $C^{(3)}_m(N;t)$.
Exact results for the $\, C^{(4)}_j$'s, based on the Wronskian cancellation
method of Sec. \ref{wronsk}, and {\em valid for any value of $\, N$}
 are displayed in \ref{C4j}. These are exact results for the palindromic
polynomials in terms of $\, F_N$ and $\, u_2(N)$, namely two {\em hypergeometric
functions associated with elliptic curves}. Thus, these analytical 
results can also be seen as an illustration
of the canonical link between our palindromic polynomials and the 
theory of elliptic curves. They  confirm the deep relation 
we find, algebraically and analytically, 
on these structures with the theory of elliptic curves. 
In a forthcoming publication we will show that the relation 
is in fact, more specifically, a close relation 
with {\em modular forms}.

\section{Conclusions}
\label{conclu}

In this paper we have proven the factorization,
 {\em for all $N$}, of the diagonal form
factor $\,f^{(n)}_{N,N}(t)$ for $n=2,\, 3$ previously seen in~\cite{mccoy1}
for $N \leq 4$ and provided a conjecture for $n=4$. 
Besides new results like the quadratic recursion (\ref{nice}),
or non trivial quartic identities (like (\ref{c33n})--
(\ref{c31n})),
one of the main result of the paper is the fact that, introducing 
the selected hypergeometric functions $\, F_N$, which are also
elliptic functions, and are simply related to the 
(simplest) form factor $\, f_{N,N}^{(1)}$, the form factors 
actually become polynomials of these $\, F_N$'s with palindromic
 polynomial coefficients. The complexity of the form factors, is,
thus, reduced to some encoding in terms of palindromic
 polynomials. As a consequence, understanding  the form factors
amounts to describing and understanding an infinite set of palindromic
 polynomials, canonically associated with elliptic curves.

 We also observe that all of these palindromic polynomials are built
from the solutions of the operator $O_2(N)$, and, therefore, are all
properties of the basic elliptic curve which underlies all
computations of the Ising model. There is a deep structure here which
needs to be greatly developed. The differential algebra approach of 
the linear differential operators associated with these palindromic polynomials
is found to be a surprisingly rich structure
canonically associated with elliptic curves. In a forthcoming
publication, we will show that such rich structures
are {\em  closely related to modular forms}.

Analytically, the conjecture and the  Wronskian method of logarithm
cancellation can be extended to large values of $n$, but the method of
proof by differential equations becomes prohibitively cumbersome for
$n \ge \, 4$. This is very similar to the situation which
 occurred for the factorization of
correlations in the XXZ model where the factorizations
 of~\cite{bk}-\cite{ksts} done for small values of the separation of the
spins by means of explicit computations on integrals was proven for
all separations in~\cite{bjmst} by means of the qKZ equation satisfied
by the correlations and not by the explicit integrals which are the
solution of this equation. This suggests that our palindromic
polynomials may profitably be considered as a specialization of 
polynomials of $n$ variables. Moreover, if the two conjectures
presented in the introduction are indeed correct, then such kind of 
structures could also have relevance to the 8 vertex model and 
to the higher genus curves which arise in the chiral Potts model. Consequently the
computations presented here could be a special case of a much larger
modularity phenomenon. This could presumably generalize the relations 
which the Ising model has with modular forms and Calabi-Yau
 structures~\cite{CalabiYauIsing}.  

%\newpage
\vspace{.1in}

{\bf Acknowledgment}

This work was supported in part by the National Science Foundation
grant PHY-0969739.

\vspace{.1in}

 \appendix

\section{Form factors in the basis $F_N$ and $F_{N+1}$}
\label{ffbasis}

By use of the contiguous relations for hypergeometric functions the
examples given in~\cite{mccoy1} of $f^{(n)}_{N,N}(t)$ 
 expressed in terms of the elliptic integrals $K(t^{1/2})$ and 
$E(t^{1/2})$ may be
re-expressed in terms of the functions $F_N$ and $F_{N+1}$. Several
examples are as follows
\begin{align}
f^{(2)}_{0,0}\,\, &=\,\,\,  \frac{t}{4} \cdot F_0\cdot F_1, 
\\
f^{(2)}_{1,1}\,\, &=\,\, \, \frac{1}{2}\, \, 
-\frac{1}{4}\, 
\, (t+1)  \left(2\,{t}^{2}+t+2 \right)\cdot  F_1^2
\nonumber\\
&\qquad \qquad 
+{\frac {3^2}{2^5}} \cdot t \cdot (4t^2+5\,t+4) \cdot F_1 \cdot  F_2 
-\frac {3^4}{2^7}\,{t}^{2} \, (t+1) \cdot   F_2^2, 
\\
f^{(2)}_{2,2}\,\, &=\, \,\,\,  1\,\, \, 
 -{\frac {1}{2^6}}\, (t+1)  
\, (64\,{t}^{4}+16\,{t}^{3}+99\,{t}^{2}+16\,t+64) \cdot F^2_2
\nonumber\\
&\qquad \quad +{\frac {5^2}{2^8\cdot 3}}
\cdot t \cdot (64t^4+88\,t^3
+105\,{t}^{2}+88\,{t} +64) \cdot  F_2 \cdot F_3
\nonumber\\
&\qquad \quad -{\frac {5^4}{2^7\cdot 3^2}} \cdot {t}^{2} \cdot  (t+1)  
\, (2\,{t}^{2}+t+2)\cdot F^2_3, 
\\
f^{(2)}_{3,3}\,\, &=\,\,\,  \frac{3}{2} 
\nonumber\\
&\quad
-{\frac {1}{2^7\cdot 3}}\cdot  (t+1) 
 \left( 576\,{t}^{6}+96\,{t}^{5}+730\,{t}^{4}
+425\,{t}^{3}+730\,{t}^{2}+96\,t+576\right)\cdot F^2_3
\nonumber\\
&\quad +{\frac {7^2}{2^{12}\cdot 3}} \cdot t \, (768t^6+928\,t^5
+1240\,{t}^{4} +1455\,{t}^{3}+1240\,{t}^{2}+928\,{t}+768)
\cdot F_3\,F_4
\nonumber\\
&\quad -{\frac {7^4}{2^{15}\cdot 3}} \cdot {t}^{2} \, (t+1)  \, (64\,{t}^{4}
+16\,{t}^{3}+99\,{t}^{2}+16\,t+64) \cdot  F_4^2.
\end{align}

For $f^{(3)}_{N,N}$ with $N=\,0,\cdots,4$
\begin{align}
\label{f30}
f^{(3)}_{0,0}\, \,&=\,\,\,
 \frac{1}{2\cdot 3}\cdot f^{(1)}_{0,0}\, \,
-\frac{1}{2\cdot 3} \, (1+t)\cdot {F_0}^{3}\,
 +\frac{1}{2^2}t\cdot {F_0}^{2} \cdot F_1,  \\
\frac{f^{(3)}_{1,1}}{t^{1/2}}
\, \,&=\,\, \, \frac{2}{3} \cdot \frac{f^{(1)}_{1,1}}{t^{1/2}} \,\,\,
 -\frac{1}{2^3\cdot 3}\left( 1+t \right)  
\left( 2^3t^2+13\,t+2^3 \right) \cdot {F_1}^{3}
\nonumber\\ 
\label{f31}
&\quad +\frac{3^2}{2^6}t \left( 8t^2+15\,t+8 \right)
 \cdot {F_1}^{2} \cdot F_2
-\frac{3^4}{2^6}t^2(t+1)F_1F^2_2
+\frac{3^5}{2^9}{t}^{3}\cdot {F_2}^{3}, \\
\frac{f^{(3)}_{2,2}}{t} \,\,  
&=\, \, \frac{7}{2\cdot 3} \cdot \frac{f^{(1)}_{2,2}}{t}
\nonumber\\
&\quad -\frac{1}{2^{10}\cdot 3}
\left( 1+t \right)  \left( 2^6\cdot 3 \cdot 7\,{t}^{4}
+1136\,{t}^{3}+3229\,{t}^{2}
+1136\,t+1344 \right)\cdot  {F_2}^{3}
\nonumber\\
&\quad +\frac{5^2}{2^{11}\cdot 3}\,t \left( 2^5\cdot 3^2t^4+596\,t^3+859
\,{t}^{2}+596\,{t}+2^5\cdot 3^2 \right)\cdot  {F_2}^{2} \cdot F_3
\nonumber\\
&\quad-\frac{5^5}{2^{10}\cdot 3^2}(t+1)(3t^2+4t+3)\, t^2 \cdot F_2 \cdot F_3^2
 \, +\frac{5^6}{2^{11}\cdot 3^4}{t}^{3} 
\left( 3t^2+8\,t+3 \right) \cdot  {F_3}^{3},
\end{align}
\begin{align}
\frac{f^{(3)}_{3,3}}{t^{3/2}} \, \,
&=\, \,\, \frac{5}{3} \cdot \frac{f^{(1)}_{3,3}}{t^{3/2}} 
\nonumber\\ 
&\quad -\frac{1}{2^{11}\cdot 3^4}( t+1)   
(2^7\cdot 3^3\cdot 5^2{t}^{6}+49680\,{t}^{5}+153306\,{t}^{4}
+160427\,{t}^{3}
\nonumber\\
&\quad ~~~~~+153306\,
{t}^{2}+49680\,t+2^7\cdot 3^3\cdot 5^2 ) \cdot  F_3^{3}
\nonumber\\
&\quad +\frac{7^2}{2^{16}\cdot 3^3}t( 2^{10}\cdot 3^2\cdot 5{t}^{6}
+79200\,{t}^{5}+128104\,{t}^{4}+168593\,{t}^{3}
\nonumber\\
&\quad ~~~~~+128104\,{t}^{2}+79200\,t+2^{10}\cdot 3^2\cdot 5)
\cdot F_3^{2}\cdot F_4
\nonumber\\
&\quad -\frac{7^4}{2^{16}\cdot 3^3}( t+1)   
t^2(2^4\cdot 3^2\cdot 5{t}^{4}+670\,{t}^{3}+1763\,{t}^{2}+670\,t+2^4\cdot
3^2\cdot 5)\cdot  F_3 \cdot F_4^{2}
\nonumber\\
&\quad +\frac{7^6}{2^{21}\cdot 3^4}t^3(2^6\cdot 5{t}^{4}+740\,{t}^{3}
+1407\,{t}^{2}+740\,t+2^6\cdot 5)\cdot  F_4^{3}, \\
\frac{f^{(3)}_{4,4}}{t^2}  \, 
&=\, \,\,  \, \frac{13}{6}\cdot \frac{f^{(1)}_{4,4}}{t^2}
\nonumber\\
&\quad-\frac{1}{2^{22}\cdot 3}\left( t+1 \right)
 (2^{14}\cdot 5 \cdot
7 \cdot 13{t}^{8}+3254272\,{t}^{7}\, 
+11474624\,{t}^{6}+8672032\,{t}^{5}
\nonumber\\
&\quad+20423231\,{t}^{4}+8672032\,{t}^{3}+
11474624\,{t}^{2}+3254272\,t+2^{14}\cdot 5\cdot 7\cdot 13) \cdot 
F_4^{3}\nonumber\\
&\quad+\frac{3^3}{2^{23}\cdot 5}t( 2^{12}\cdot 3\cdot 5^2\cdot
 7{t}^{8}+3334912\,{t}^{7}+4845120\,{t}^{6}+7068720\,{t}^{5}
\nonumber\\
&\quad+8865649\,{t}^{4}+7068720\,{t}^{3}+
4845120\,{t}^{2}+3334912\,t+2^{12}\cdot 3\cdot 5^2 \cdot 7)
\cdot F_4^{2}\cdot F_5
\nonumber\\
&\quad-\frac{3^6}{2^{20}\cdot 5^3}t^2(t+1)
( 2^4\cdot 3^2\cdot5\cdot 7^2{t}^{6}+26292\,{t}^{5}+69377\,{t}^4
+78580\,{t}^{3}
\nonumber\\
&\quad+69377\,{t}^{2}+26292\,t+2^4\cdot 3^2\cdot 5\cdot 7^2) \cdot 
F_4 \cdot  F_5^{2}
\nonumber\\
&\quad+ \frac{3^8}{2^{20}\cdot 5^3}t^3( 2^3\cdot 3^3 \cdot 5 \cdot
7\,{t}^{6}
+2^3\cdot 3^3\cdot7\cdot
11{t}^{5}+28413\,{t}^{4}+46432\,{t}^3
\nonumber\\
\label{f34}
&\quad+28413\,{t}^{2}+2^3\cdot 3^3\cdot 7\cdot 11\,t+2^3\cdot 3^3\cdot
5\cdot 7) \cdot  F_5^{3}.
\end{align}
The coefficients which are not given in factored form all contain
large prime factors.

For $f^{(4)}_{N,N}$ with $N\,=\,\, 0,1,2,3$
\begin{align}
f^{(4)}_{0,0}\,\,&=\,\,\, \, \frac{1}{3}\cdot f^{(2)}_{0,0} 
-\frac{1}{2^2\cdot 3}\cdot t\cdot F_0^4\, \,
 +\frac{1}{2^5} \cdot  t \cdot F_0^2 \cdot F_1^2,\label{f400} \\ 
f^{(4)}_{1,1}\,\,
&=\,\,\,\,  -\frac{1}{2^3\cdot 3} + \frac{5}{2\cdot 3}\cdot f^{(2)}_{1,1}
+\frac{1}{2^5\cdot 3}(4t^4+4t^3+15t^2+4t+4)(t+1)^2 \cdot F_1^4
\nonumber\\
&\quad-\frac{3}{2^7}t(t+1)(8t^4+18t^3+35t^2+18t+8)\cdot F_1^3 \cdot F_2
\nonumber\\
&\quad+\frac{3^4}{2^{11}}\cdot t^2\, 
(8t^4+28t^3+45t^2+28t+8)\cdot F_1^2 \cdot F_2^2
\nonumber\\
&\quad-\frac{3^5}{2^{12}} \cdot 
t^3  \cdot  (t+1) \cdot (4t^2+11t+4)\cdot F_1 \cdot F_2^3 
+\frac{3^7}{2^{15}}t^4(t^2+4t+1) \cdot F_2^4, \label{f411}
\end{align}
\begin{align}
f^{(4)}_{2,2}\,\, 
&= \,\,\, \,-\frac{1}{3}\,\, + \frac{2^2}{3}\cdot f^{(2)}_{2,2}
\nonumber\\
&\quad +{\frac {1}{2^{14}\cdot 3}}\,
 \left( 2^{14}\,{t}^{10}+40960\,{t}^{9}
+84480\,{t}^{8}+136640\,{t}^{7}+176180\,{t}^{6}\right.
\nonumber\\
&\quad \left.+201075\,{t}^{5}+176180\,{t}^{4}+136640\,{t}^{3}
+84480\,{t}^{2}+40960\,t+2^{14} \right) \cdot F_2^{4}
\nonumber\\
&\quad -{\frac {5^2}{2^{14}\cdot 3^2}}\, t\left( t+1 \right) 
 \left( 2^{13}\,{t}^{8}
+13312\,{t}^{7}+29504\,{t}^{6}+36320\,{t}^{5}+45337\,{t}^{4}\right.
\nonumber\\
&\quad \left.+36320\,{t}^{3}+29504\,{t}^{2}+13312\,t
+2^{13} \right)\cdot  F_2^{3} \cdot F_3
\nonumber\\
&\quad +{\frac {5^4}{2^{17}\cdot 3^2}}\, {t}^{2}\left( 2^{12}\,{t}^{8}
+11264\,{t}^{7}+21760\,{t}^{6}+31576\,{t}^{5}+36209\,{t}^{4}\right.
\nonumber\\
&\quad \left.+31576\,{t}^{3}+21760\,{t}^{2}+11264\,t
+2^{12} \right)\cdot  F_2^{2} \cdot F_3^{2}
\nonumber\\
&\quad -{\frac {5^6}{2^{15}\cdot 3^4}}\cdot  {t}^{3}\, (t+1)  
\left( 2^8\,{t}^{6}+480
\,{t}^{5}+906\,{t}^{4}+979\,{t}^{3}+906\,{t}^{2}\right.\nonumber\\
&\quad \left.+480\,t+2^8 \right) \cdot F_2 \cdot F_3^{3}
\nonumber\\
&\quad +{\frac {5^8}{2^{15}\cdot 3^4}}\, {t}^{4}\left( 2^5\,{t}^{6}
+96\,{t}^{5}+177\,{t}^{4}
+224\,{t}^{3}+177\,{t}^{2}+96\,t+2^5 \right) \cdot
F_3^{4},\label{f422}
\end{align}
\begin{align}
f^{(4)}_{3,3} \,\, 
&= \, \, \, \,	-\frac{7}{2^3} + \frac{11}{2\cdot 3}\cdot f^{(2)}_{3,3}
\nonumber\\
&\quad+\frac {1}{2^{16}\cdot 3^4}\, \left( 2^{13}\cdot 3^4\cdot 7{t}^{14}
+10838016\,{t}^{13}+19643904\,{t}^{12}+34169856\,{t}^{11}\right.
\nonumber\\
&\quad\left.+50403584\,{t}^{10}+62791680\,{t}^{9}+73309425
\,{t}^{8}+79935700\,{t}^{7}+73309425\,{t}^{6}\right.
\nonumber\\
&\quad\left.+62791680\,{t}^{5}+50403584\,{t}^{4}+34169856\,{t}^{3}
+19643904\,{t}^{2}+10838016\,t\right.\nonumber\\
&\quad\left.+2^{13}\cdot 3^4\cdot 7 \right) \cdot F_3^{4}
\nonumber\\
&\quad-{\frac {7^2}{2^{19}\cdot 3^4}}\, t\left( t+1 \right)  
\left( 2^{14}\cdot 3^3\cdot 7{t}^{12}+4257792\,{t}^{11}+9547776\,{t}^{10}
+13813120\,{t}^{9}\right.
\nonumber\\
&\quad\left.+19341120\,{t}^{8}+21399090\,{t}^{7}+24976435\,{t}^{6}
+21399090\,{t}^{5}+19341120\,{t}^{4}\right.
\nonumber\\
&\quad\left.+13813120\,{t}^{3}+9547776\,{t}^{2}+4257792\,t
+2^{14}\cdot 3^3\cdot 7 \right)\cdot  F_3^{3} \cdot F_4
\nonumber\\
&\quad+{\frac {7^4}{2^{25}\cdot 3^5}}\, {t}^{2} \cdot 
\left( 2^{15}\cdot 3^2\cdot 7\, \, {t}^{12}+4988928\,{t}^{11}
+9680384\,{t}^{10}+15992320\,{t}^{9} \right.
\nonumber\\
&\quad\left.+21863120\,{t}^{8}+26325960\,{t}^{7}+28527015\,{t}^{6}
+26325960\,{t}^{5} +21863120\,{t}^{4}\right.
\nonumber\\
&\quad\left.+15992320\,{t}^{3}+9680384\,{t}^{2} +4988928\,t
+2^{15}\cdot 3^2\cdot 7 \right) \cdot 
F_3^{2} \cdot F_4^{2}
\nonumber\\
&\quad-{\frac {7^6}{2^{27}\cdot 3^4}}\, {t}^{3} \cdot \left( 
t+1 \right)  \left( 2^{14}\cdot 3\cdot 7{t}^{10}
+501760\,{t}^{9}+1191680\,{t}^{8} +1548640\,{t}^{7}\right.
\nonumber\\
&\quad\left.+2065400\,{t}^{6}+2169745\,{t}^{5}+2065400\,{t}^{4}+
1548640\,{t}^{3}\right.
\nonumber\\
&\quad\left.+1191680\,{t}^{2}+501760\,t
+2^{14}\cdot 3\cdot 7 \right) \cdot  F_3 \cdot F_4^{3}
\nonumber\\
&\quad+{\frac {7^8}{2^{31}\cdot 3^4}}\, {t}^{4}\cdot 
\left( 2^{12}\cdot 7{t}^{10}
+71680\,{t}^{9}+147840\,{t}^{8}+235040\,{t}^{7}
+299555\,{t}^{6}\right.
\nonumber\\
&\quad\left.+339180\,{t}^{5}+
299555\,{t}^{4}+235040\,{t}^{3}+147840\,{t}^{2}+71680\,t+2^{12}\cdot 7
 \right) \cdot F_4^{4}.\label{f433} 
\end{align}

For $f^{(5)}_{N,N}$ with $N\, =\,\,1,2,3$
\begin{align}
\frac{f_{1,1}^{(5)}}{t^{1/2}}\,\, \, &=\, \, \,  
-\frac{2^2}{5} \cdot \frac{f^{(1)}_{1,1}}{t^{1/2}} 
+ \frac{f^{(3)}_{1,1}}{t^{1/2}}
\nonumber\\
&\quad +\frac{1}{2^6\cdot 3\cdot 5}
(t+1)^2(2^6+136t^3+159t^2+136t+2^6) \cdot F_1^5
\nonumber\\
&\quad -\frac{3}{2^8}t(t+1)(2^5t^4+80t^3+99t^2+80t+2^5)\cdot F_1^4 \cdot F_2
\nonumber\\
&\quad +\frac{3^3}{2^{12}}t^2(2^7t^4+368t^3+483t^2+368t+2^7) \cdot F_1^3 \cdot F_2^2
\nonumber\\
&\quad -\frac{3^5}{2^{10}}t^3(t+1)(4t^2+5t+4)\cdot F_1^2 \cdot F_2^3
\nonumber\\
&\quad +\frac{3^7}{2^{15}}t^4(8t^2+13t+8)\cdot F_1\cdot F_2^4
\, \, -\frac{3^9}{2^{15}\cdot 5}(t+1)\cdot t^5\cdot F_2^5, 
\end{align}
\begin{align}
\frac{f^{(5)}_{2,2}}{t}\,  \, \, &= \, \, \,  \,
-\frac{137}{2^3\cdot5} \cdot \frac{f^{(1)}_{2,2}}{t} 
+ \frac{3}{2}\cdot\frac{f^{(3)}_{2,2}}{t}
\nonumber\\
&\quad+{\frac {1}{2^{18}\cdot 3\cdot 5}}\, 
\left( 8\,{t}^{2}+7\,t+8 \right)  
\left(2^{9}\cdot 3\cdot 61\,{t}^{8}+241856\,{t}^{7}+508200\,{t}^{6}
+708609\,{t}^{5}\right.
\nonumber\\
&\quad\left.+780244\,{t}^{4}+708609\,{t}^{3}+508200\,{t}^{2}
+241856\,t+2^{9}\cdot 3\cdot 61 \right) \cdot  F_2^{5}
\nonumber\\
&\quad-{\frac {5^2}{2^{18}\cdot 3^2}}\, t\, \left( t+1 \right) 
 \left( 92160\,{t}^{8} +239360\,{t}^{7}+540576\,{t}^{6}+723924\,{t}^{5}
\right.
\nonumber\\
&\quad\left.+868861\,{t}^{4} +723924\,{t}^{3}
+540576\,{t}^{2}+239360\,t+92160 \right) 
\cdot  F_2^{4} \cdot F_3
\nonumber\\
&\quad+{\frac {5^4}{2^{20} \cdot 3^3}}\, {t}^{2}\left( 90624\,{t}^{8}
+338816\,{t}^{7}+743304
\,{t}^{6}+1122432\,{t}^{5}+1278697\,{t}^{4}\right.
\nonumber\\
&\quad\left.+1122432\,{t}^{3}+743304\,{t}^{2}
+338816\,t+90624 \right) \cdot {F_2}^{3}\, {F_3}^{2}
\nonumber\\
&\quad-{\frac {5^6}{2^{17}\cdot 3^4}}\, {t}^{3}\left( t+1 \right)  
\left( 1392\,{t}^{6}+4010\,{t}^{5}+6983\,{t}^{4}
+8136\,{t}^{3}\right.
\nonumber\\
&\quad\left.+6983\,{t}^{2}+4010\,t+1392 \right) \cdot F_2^{2}\cdot F_3^{3}
\nonumber\\
&\quad+{\frac {5^8}{2^{20}\cdot 3^5}}\, {t}^{4}
\left( 684\,{t}^{6}+2752\,{t}^{5}+5161\,{t}^{4}
+6240\,{t}^{3}+5161\,{t}^{2}+2752\,t\right.\nonumber\\
&\quad\left.+684 \right)\cdot  F_2 \cdot F_3^{4}
\nonumber\\
&\quad-{\frac {5^9}{2^{19}\cdot 3^6}}\, {t}^{5}
\, (t+1)  \left( 42\,{t}^{4}
+133\,{t}^{3}+167\,{t}^{2}+133\,t+42 \right) \cdot  F_3^{5},  
\end{align}
\begin{align}
 \frac{f^{(5)}_{3,3}}{t^{3/2}} 
\, \, \, &=  \, \, \,\, 
-\frac{127}{3\cdot 5} \cdot \frac{f^{(1)}_{3,3}}{t^{3/2}} 
+ 2\cdot\frac{f^{(3)}_{3,3}}{t^{3/2}}
\nonumber\\
&\quad+{\frac {1}{2^{20}\cdot 3^5\cdot 5}}\left( 2^{16}\cdot 3^4\cdot
  5\cdot 17\,{t}^{14}
+1377976320\,{t}^{13}+3016452096\,{t}^{12}\right.
\nonumber\\
&\quad\left.+5930641920\,{t}^{11}+9308313280\,{t}^{10}
+12328157240\,{t}^{9}+14834544515\,{t}^{8}\right.
\nonumber\\
&\quad\left.+15849843292\,{t}^{7}+14834544515\,{t}^{6}
+12328157240\,{t}^{5}+9308313280\,{t}^{4}\right.
\nonumber\\
&\quad\left.+5930641920\,{t}^{3}+3016452096\,{t}^{2}+1377976320\,t
+2^{16}\cdot 3^4\cdot 5\cdot 17 \right) \cdot  F_3^{5}
\nonumber\\
&\quad-{\frac {7^2}{2^{22}\cdot 3^5}}\, {t}\, (t+1)  
\left( 2^{19}\cdot 3^3\cdot 5\,{t}^{12}
+151511040\,{t}^{11}+351000576\,{t}^{10}\right.
\nonumber\\
&\quad\left.+605214208\,{t}^{9}+835692208\,{t}^{8}
 +1025976166\,{t}^{7}  +1112168875\,{t}^{6}
\right.
\nonumber\\
&\quad\left.+1025976166\,{t}^{5} +835692208\,{t}^{4}+605214208\,{t}^{3}
+351000576\ {t}^{2}\right.\nonumber\\
&\quad\left.+151511040\,t+2^{19}\cdot 3^3\cdot 5 \right) 
\cdot F_3^{4} \cdot F_4
\nonumber\\
&\quad+{\frac {7^4}{2^{29}\cdot 3^5}}\, {t}^{2}
\left( 2^{18}\cdot 3^3\cdot 5^2\,{t}^{12}+
572129280\,{t}^{11}+1334317056\,{t}^{10}\right.\nonumber\\
&\quad\left.+2446757888\,{t}^{9}+3545541888\,{t}^{8}+4425343776\,{t}^{7}
+4784608975\,{t}^{6}\right.\nonumber\\
&\quad\left.+4425343776\,{t}^{5}+3545541888\,{t}^{4}+2446757888\,{t}^{3} 
+1334317056\,{t}^{2}\right.\nonumber\\
&\quad\left.+572129280\,t+2^{18}\cdot 3^3\cdot 5^2 \right) 
\cdot F_3^{3}\cdot F_4^{2}
\nonumber\\
&\quad-{\frac {7^6}{2^{27}\cdot 3^5}}\, {t}^{3}\left( t+1 \right)  
\left( 2^{13}\cdot 3 \cdot 5\cdot 7\,{t}^{10}+2007040\,{t}^{9}
+4885888\,{t}^{8} +7228048\,{t}^{7}\right.
\nonumber\\
&\quad\left.+9666130\,{t}^{6}+10423545\,{t}^{5}+9666130\,{t}^{4}
+7228048\,{t}^{3}+4885888\,{t}^{2}\right.
\nonumber\\
&\quad\left.+2007040\,t+2^{13}\cdot 3\cdot 5 \cdot 7 \right) 
 \cdot F_3^{2} \cdot F_4^{3}
\nonumber\\
&\quad+{\frac {7^8}{2^{34}\cdot 3^5}}\, {t}^{4}\left( 2^{14}\cdot 5 \cdot 13
\,{t}^{10}+3665920\,{t}^{9}+9078784\,{t}^{8}
+15185664\,{t}^{7}\right.
\nonumber\\
&\quad\left.+20375540\,{t}^{6}+22605185\,{t}^{5}
+20375540\,{t}^{4}+15185664\,{t}^{3}+9078784\,{t}^{2}\right.
\nonumber\\
&\quad\left.+3665920\,t+2^{14}\cdot 5 \cdot 13 \right) \cdot  F_3  \cdot F_4^{4}
\nonumber\\
&\quad-\frac {7^{10}}{2^{35}\cdot 3^5\cdot 5}\, {t}^{5}\left( t+1 \right)  \left( 
2^{13}\cdot 5\,{t}^{8}+104960\,{t}^{7}+267136\,{t}^{6}+319904\,{t}^{5}\right.
\nonumber\\
&\quad\left.+436441\,{t}^{4}+319904\,{t}^{3}+267136\,{t}^{2}
+104960\,t+2^{13}\cdot 5 \right) \cdot F_4^{5}.  
\end{align}

\section{Polynomial solution calculations for $C^{(2)}_1(N;t)$}
\label{polsol1ap}

We here give explicitly the calculational details for $C^{(2)}_1(N;t)$.  

Using the form (\ref{initial}) in the inhomogeneous equation
(\ref{c1ode}) we find the recursion relation for the coefficients
$c^{(2)}_{1;n}$ for $n\neq N,~N+1,~N+2,~N+3$
\begin{align}
\label{c1rr1}
&2n \cdot  (n-N)\,(n-2N-1) \cdot c^{(2)}_{1;n}
\nonumber\\
&\quad -\{2n^3-6Nn^2-2(4+N-2N^2)n+5+6N\} \cdot c^{(2)}_{1;n-1}
\nonumber\\
&\quad -\{2n^3-6(3+N)n^2+(46+34N+4N^2)n\, -35-38N-8N^2\} \cdot c^{(2)}_{1;n-2}
\nonumber\\
&\quad +2\, (n-2)(n-2N-3)(n-N-3)\cdot c^{(2)}_{1;n-3}
\, \, =\,\, \,  0.
\end{align}
where by definition $c^{(2)}_{1;n}=0$ for $n\leq -1$.
This recursion relation
has four terms instead of the three terms in the corresponding
relation (\ref{rr2}) for $c^{(2)}_{2;n}$.
We note that, if we send $n\, \rightarrow \,  2N-n+3$
in (\ref{c1rr1}),  
we see that $c^{(2)}_{1;n}$ and $c^{(2)}_{2N-n}$ satisfy the same
equation. 
Since the coefficient of $c^{(2)}_{1;n}$ vanishes for $n=\,0$, the term
$c^{(2)}_{1;0}$ is not determined from (\ref{c1rr1}) and by convention
we set $c^{(2)}_{1;0}=1$

Following the procedure used for $C^{(2)}_2(N;t)$ we note that equation
(\ref{c2ode}) will be satisfied to order $t^N$ if we choose the 
$c^{(2)}_{1;n}$ for $0 \, \leq\,  n \, \leq\,  N-1$ to be the corresponding 
coefficients in 
$ \, t^{-1} \cdot u_2(N;t)\cdot u_2(N+1;t)$ and hence (\ref{dn1}) follows.

The inhomogeneous recursion relations for $n\,=\, N,\,\,N+1$ are
\begin{multline}
\label{c1rr2}
A^{(2)}_1\{-(2N^2+2N-5) \cdot c^{(2)}_{1;N-1}\, \,  \, 
+(8N^2+8N-35) \cdot c^{(2)}_{1;N-2}
\\ 
-6(N-2)(N+3) \cdot c^{(2)}_{1;N-3}\}
\,\, \, \,  =\, \, \, \, \, \, 
-2N^2(2N+1)^2\,\lambda_N^2,
\end{multline}
\begin{multline}
\label{c1rr3}
A^{(2)}_{1}\{-2\,N\,(N+1)\cdot c^{(2)}_{1;N+1}\, \,  -(2N+1)^2 \cdot c^{(2)}_{1;N}\, 
\\
 +(6N^2+6N-5) \cdot c^{(2)}_{1;N-1} \, 
+4 \, (N+2)(N-1) \cdot c^{(2)}_{1;N-2}\}\, \, 
\\
 \, \, =\,\,\,
-\frac{(2N+1)^2 \,(4N^3+4N^2-4N-1)\,\lambda_N^2}
{4 \, (N+1)},
\end{multline}
and the relations for $N+2,~N+3$ are identical with 
$N,~N+1$, respectively, with the (palindromic) replacement 
\begin{equation}
c^{(2)}_{1;N-m}\, \, \, \longrightarrow\, \, \,  \, \, \, c^{(2)}_{1;N+m}.
\end{equation}

If there were no inhomogeneous term (\ref{c1rr2}) would be a new
constraint in the coefficients $c^{(2)}_{1;n}$ for $n=N-1,~N-2,~N-3$. 
However this constraint
does not hold (because the solution to the homogeneous equation has a
term $t^{N+1}\ln t$). 

The normalizing constant $A^{(2)}_1$ can be evaluated from
(\ref{c1rr2}) and
the sum on the LHS of (\ref{c1rr2}) is evaluated the same way the
corresponding sum was for $C^{(2)}_2(N;t)$, by comparing with the full
solution $t^{-1}\,u_2(N;t)\,u_2(N+1;t)$ of the homogeneous equation.
Thus we find
\begin{multline}
 -(2N^2+2N-5)\cdot \, c^{(2)}_{1;N-1}(N)\, \,
+(8N^2+8N-35) \cdot \, c^{(2)}_{1;N-2}(N)
\\
-6\,(N+3)\,(N-2)\cdot \, c^{(2)}_{1;N-3}(N)  \, \,  \,  
=\, \, \,  \,  -2\,N^2 \, (N+1)\,\lambda_N^2, 
\end{multline}
and, hence, we find from (\ref{c1rr2}) 
\begin{equation}
A^{(2)}_1  \,  \, = \, \, \, \, N\beta_N.
\end{equation}

It remains to compute $c^{(2)}_{1;N}$ from (\ref{c1rr3}).
We obtain the palindromic solution by requiring that
$c^{(2)}_{1,N+1}\, =\,\, c^{(2)}_{1;N-1}$ and thus 
(\ref{c1rr3}) reduces to
\begin{multline}
\label{method1}
(2N+1)^2 \cdot c^{(2)}_{1;N}\, \,
  +(4N^2+4N-5) \cdot c^{(2)}_{1;N-1}\,
+4(N+2)(N-1)\cdot c^{(2)}_{1;N-2}
\\
\qquad = \, \, \,\,\,   -\frac{(2N+1)^2 \, (4N^3+4N^2-4N-1)\,\lambda_N^2}
{4(N+1)}.
\end{multline}
An equivalent and more efficient method for evaluating $c^{(2)}_{1;N}$,
which avoids the need to evaluate the sums on the LHS of
(\ref{method1}), is to directly evaluate $C^{(2)}_1(N;t)$ in terms of
$C^{(2)}_2(N;t)$ by use of the coupled equation (\ref{eqn3}). From
this we find
\begin{equation}
 \label{morec21n}
c^{(2)}_{1;N}(N) \, \,\,  = \, \,  \,  \, \,
 \, \lambda_N^2 \, \, \,  \, 
+\sum_{k=0}^{N-1} \, a_k(N)\cdot a_{N-1-k}(N), 
\end{equation}
and, by explicitly evaluating the  sum in (\ref{morec21n}), we obtain
the result (\ref{c21n}). Finally, the $c^{(2)}_{1;\,n}$ for
$N+1\leq n\leq 2N$ are determined from the palindromy of (\ref{c1rr1}).

\section{Coupled differential equations for $C^{(3)}_m(N;t)$}
\label{appcoupled}

The four coupled differential equations for $C^{(3)}_m(N;t)$ are
\begin{align}
&- \,{\frac {  2\,N+1  }{
 \, 2\cdot  (t-1) \, t}} \cdot C^{(3)}_0(t)\, \,  \, 
-{\frac { \, (N+1)  
\left( 2\,tN+1+t \right) }{ t^{2} \cdot (2\,N+1) 
 \left( t-1 \right) }} \cdot C^{(3)}_1 (t) 
\nonumber\\
& -2\,{\frac { (N+1)^{2} \, (2\,tN+3) }{{t}^{2} \, (2\,N+1) ^{2}
 \left( t-1 \right) }} \cdot C^{(3)}_2(t) \, 
\, \, -8\,{\frac { (N+1)^{3} \, (tN-t+3) }{ (t-1) 
 \, (2\,N+1)^{3}\, {t}^{2}}} \cdot  C^{(3)}_3(t)
\nonumber\\
&+{\frac {  tN+2\,t-N-1  }{
 \, (t-1) t}} \cdot {\frac {d}{dt}}{C^{(3)}_0}(t)  \, 
+2\,{\frac { (N+1) 
 \, (tN-N+t) }{t \, (t-1)  \, (2\,N+1) }}
 \cdot {\frac {d}{dt}}C^{(3)}_1(t) 
\nonumber \\
 & +4\,{\frac {(N+1)^{2}
 \, (tN-N+1) }{t \, (t-1)  \, (2\,N+1)^2}} \cdot
 {\frac {d}{dt}}C^{(3)}_2 (t)
+8\,{\frac { (N+1)^{3} \, (tN-N+2-t)  }{t \, (2\,N+1)^{3} \, (t-1) }}
\cdot {\frac {d}{dt}}C^{(3)}_3(t)\, 
\nonumber\\
&+{\frac {d^{2}}{d{t}^{2}}}C^{(3)}_0 (t)\, 
 +2\,{\frac {N+1 }{2\,N+1}} \cdot {\frac {d^{2}}{d{t}^{2}}}C^{(3)}_1(t) 
\label{coupled1}
+4\,{\frac { (N+1)^{2} }{ (2\,N+1)^{2}}} 
\cdot {\frac {d^{2}}{d{t}^{2}}}C^{(3)}_2 (t)\nonumber\\
&+8\,{\frac { \left( N+1 \right)^{3} 
 }{ \left( 2\,N+1 \right)^{3}}} \cdot
 {\frac {d^{2}}{d{t}^{2}}} C^{(3)}_3(t) 
\,\, \, =\,\,  \, \, \,\, \frac{3}{4}\,\,{t}^{N-1}
 \cdot (2\,N+1)\cdot  B_0(N),
\end{align}
\begin{align}
&-2\,{\frac { (N+1)  \, (2\,N+6\,tN+3\,t+2) }
{{t}^{2} \, (2\,N+1)^{2}}} 
\cdot C^{(3)}_1(t) \nonumber\\
& -8\,{\frac {(N+1)^{2} 
\, (4\,tN+4\,N+5+t) }{{t}^{2} \, (2\,N+1)^{3}}} \cdot C^{(3)}_2(t) 
\nonumber\\
 &-24\,{\frac {
 (N+1)^{3} \, (6\,N+2\,tN+9-2\,t)}{ (2\,N+1)^4 \, t^{2}}}
\cdot C^{(3)}_3(t)\, \, \, 
+6\,{\frac {d}{dt}}C^{(3)}_0(t)
\nonumber\\
 & +4\,{\frac { (N+1)  \, (5\,tN+N+3\,t+1) }{t \, (2\,N+1)^{2}}}
\cdot   {\frac {d}{dt}}{C^{(3)}_1}(t)
\nonumber\\
& 
+8\,{\frac {  (N+1)^{2} 
\, (2\,N+4\,tN+t+4) }{t \, (2\,N+1)^{3}}} \cdot {\frac {d}{dt}}C^{(3)}_2(t) 
\nonumber\\
 &+48\,{\frac { (N+1)^{3} \, (tN+N+3-t)}{t \, (2\,N+1)^{4}}} 
\cdot  {\frac {d}{dt}}C^{(3)}_3(t)\, 
+4\,{\frac {  (t-1)  \, (N+1) }{ (2\,N+1)^2}}
\cdot  {\frac {d^{2}}{d{t}^{2}}}C^{(3)}_1(t) 
\nonumber\\
 &+16\,{\frac { (N+1)^2
  \, (t-1) }{ (2\,N+1)^3}}\cdot  {\frac {d^{2}}{d{t}^{2}}}C^{(3)}_2(t) \, 
+48\,{\frac { (N+1)^{3}
 \, (t-1)}{ (2\,N+1)^{4}}} \cdot {\frac {d^{2}}{d{t}^{2}}}C^{(3)}_3 (t)
\nonumber\\
\label{coupled2}
 &=\, \, \,\,\,   \frac{3}{2}\, \,{t}^{N-1} 
\cdot (2\,{N}^{2}t+t+4\,tN-2\,N-2\,{N}^{2}) \cdot  B_0(N), 
\end{align}

\begin{align}
&C_1^{(3)}(t)\, \,  +4\,{\frac {(N+1) \, (2\,N+2-t)  }{t \, (2\,N+1)^{2}}}
\cdot C^{(3)}_2 (t) \, \, 
 +4\,{\frac {  (t-1)
 \, (N+1)}{ (2\,N+1) ^{2}}} \cdot  {\frac {d}{dt}}C^{(3)}_2(t) 
\nonumber\\
 &+4\,{\frac { (N+1)^2 \, (12\,{N}^{2}-16\,tN-2\,{t}^{2}
N+30\,N +2\,{t}^{2}+18-17\,t)}{(2\,N+1)^{4}\, {t}^{2}}}
\cdot  C^{(3)}_3 (t) 
\nonumber\\
&+8\,{\frac { (t-1) \, (N+1)^{2} \, (tN+5\,N-t+5) }
{t \, (2\,N+1) ^{4}}} \cdot {\frac {d}{dt}}C^{(3)}_3(t)
\nonumber\\
 \label{coupled4}
 &+8\,{\frac { (t-1)^{2} \, (N+1)^{2}}{ (2\,N+1)^{4}}} 
\cdot {\frac {d^{2}}{d{t}^{2}}}C^{(3)}_3(t)
\, \, \,  \,
=\, \, \,\, \, \, 
\frac{3}{4}\, \frac{(2N+1)^2}{(N+1)} \cdot {t}^{N+1} \cdot B_0(N),
\end{align}
\begin{align}
&6\cdot C^{(3)}_0(t)  \, \, \, 
+4\,{\frac {(N+1)  \, (4\,N+2\,tN+4-t) }{t \, (2\,N+1)^2}}
\cdot C^{(3)}_1 (t) 
\nonumber\\
 &+8\,{\frac {(N+1) ^{2} 
\left( 4\,{N}^{2}+8\,{N}^{2}t -10\,{t}^{2}N+10\,tN+12\,N -t
-4\,{t}^{2}+8 \right) }{ (2\,N+1)^{4}{t}^{2}}} \cdot C^{(3)}_2 (t)
\nonumber\\
 &+48\,{\frac { (N+1)^{3} 
\, (4\,{N}^{2}-2\,{t}^{2}N-10\,tN+16\,N+2\,{t}^{2}+13
-14\,t) }{ (2\,N+1) ^{5}\, {t}^{2}}} \cdot C^{(3)}_3 (t) 
\nonumber\\
 &+16\,{\frac { (t-1)  \, (N+1) }{ (2\,N+1)^{2}}}
 \cdot {\frac {d}{dt}}{C^{(3)}_1}(t) 
\nonumber\\
 &+16\,{\frac {(t-1)  \,
(N+1)^{2} \, (5\,tN+3\,N+2\,t+3) }{t \, (2\,N+1)^4}} 
\cdot  {\frac {d}{dt}}{C^{(3)}_2}(t)  
\nonumber\\
 &+96\,{\frac { (N+1)^3 \, (t-1)  \, (tN+3\,N+4-t)
 }{t \left( 2\,N+1 \right)^{5}}} \cdot {\frac {d}{dt}}C^{(3)}_3(t)
\nonumber\\
 &+16\,{\frac { (t-1)^{2} \, (N+1) ^{2}}{ (2\,N+1) ^{4}}}
 \cdot  {\frac {d^{2}}{d{t}^{2}}}C^{(3)}_2(t) \, \, \,
+96\,{\frac { (t-1)^2 (N+1)^3 }{ (2\,N+1)^5}}
\cdot {\frac {d^{2}}{d{t}^{2}}} C^{(3)}_3(t)
\nonumber\\
\label{coupled3}
 & =\, \,\, \,\,  3\,\,{t}^{N} \cdot (3\, N \,(t-1)\, \, +2\,t-1) \cdot B_0(N),
\end{align}
where $B_0(N)$ is given by (\ref{b0def}).

\section{The ODE and recursion relation for $C^{(3)}_3(N;t)$}
\label{appc3}

The ODE for $C^{(3)}_3(N;t)$ can be found by carefully using the four coupled
ODEs (\ref{coupled1})--(\ref{coupled3}). First use 
(\ref{coupled4}) to solve for $C^{(3)}_1(N;t)$ and then use this is  
in Equation (\ref{coupled3}) in order to solve for $C^{(3)}_0(N;t)$. Next, 
use both $C^{(3)}_0(N;t)$ and $C^{(3)}_1(n;t)$ in (\ref{coupled1}) 
and  (\ref{coupled2}) to produce ODEs of orders four in $C^{(3)}_2(N,t)$
and five in $C^{(3)}_3(N;t)$ in (\ref{coupled1}) and orders three 
in $C^{(3)}_2(N;t)$ and four in $C^{(3)}_3(N;t)$ in (\ref{coupled2}).

In the new (\ref{coupled1}), the fourth derivative of $C^{(3)}_2(N;t)$ 
can be solved in terms of the other derivatives, and likewise in the
new (\ref{coupled2}), the third derivative of $C^{(3)}_2(N;t)$ can be 
solved in terms of the other derivatives. Taking the derivative of 
the expression for the third derivative of $C^{(3)}_2(N;t)$ and 
equating it to the expression for the fourth derivative of 
$C^{(3)}_2(N;t)$ we find an alternate expression for the third 
derivative of $C^{(3)}_2(N;t)$. 
Finally, equating the two expressions for the third derivative 
of $C^{(3)}_2(N;t)$, a full cancellation of all of the derivatives of 
$C^{(3)}_2(N;t)$ takes place, 
leaving a fifth order ODE in terms of only $C^{(3)}_3(t)$
\begin{align}
&4\,  \left[ 2\, \left( N-1 \right)  \left( 2\,N+1 \right) 
 \left( 3\,N+1 \right)  \left( N+1 \right) t^{4}\right.
\nonumber\\
&\qquad\left.- \, (2\,N+3)  \, 
(36\,N^{3}-7\,N^{2}-69\,N-32)\,  t^{3}\right.
\nonumber\\
&\qquad\left.
+4\, (N+2)  \, (36\,N^{3}-10\,N^{2}-116\,N-69)\, t^{2}\right.
\nonumber\\
&\qquad\left.- (2\,N+5)  \, (60\,N^{3}-23\,N^{2
}-275\,N-188) t\right.
\nonumber\\
&\qquad\left.+18\, (2\,N+3)  \, (N+3) 
 \, (N-3)  \, (N+1)  \right] \cdot C^{(3)}_3(t)
\nonumber\\
 & -8  \left[ (N-1) 
 \, (2\,N+1)  \, (3\,N+1)  \, (N+1) \, t^{4} \right.
\nonumber\\
&\qquad\left.- (-130\,N+40\,N^{3}-47+24\,N^{4}-73\,N^{2}) \, t^{3}
\right.\nonumber\\
&\qquad\left.+2\, \left( 18\,N^{4}-129+45\,N^{3}-113\,N^{2}-270\,N
 \right) t^{2}\right.
\nonumber\\
&\qquad\left.- (-253\,N^{2}-422+24\,N^{4}-740\,N+80\,N^3)\,  t\right.
\nonumber\\
&\qquad\left.+ \left( N+1 \right)  \left( 6\,N^{3}+19\,N^{2}-114\,
N-211 \right)  \right]  \cdot  t \cdot {\frac {d}{dt}}C^{(3)}_3 (t)
\nonumber\\
& +20 \, (t-1)  \left[ 2\,N \, (N-1)  \, (N+1) \cdot  t^{3} \right.
\nonumber\\
&\quad\qquad\qquad\left.-3\, (2\,N^3-4\,N^{2}
-8\,N-3)\cdot   t^2 +3\, (-13+2\,N^{3}-8\,N^{2}-24\,N)\,  t\right]
\nonumber\\
&\quad\qquad\qquad\left.-2\, (N-9)  \, (N+2) 
 \left( N+1 \right)  \right] \cdot 
  t^2 \cdot {\frac {d^{2}}{dt^{2}}}C^{(3)}_3 (t)
\nonumber\\
 & +40 \, (t-1)^2  \left[ 
 (N-1)^2 \, t^{2}\, - \, (4\,N+1+2\,N^{2}) t
+ \left( N+5 \right)  \left( N+1 \right)  \right] \cdot  t^3 \cdot 
{\frac {d^{3}}{dt^{3}}}C^{(3)}_3 (t)
\nonumber\\
& -40 \, (t-1)^{3} \, 
  \left[ (N-1) t-N-1 \right]\,  t^4 \cdot 
{\frac {d^{4}}{dt^{4}}}C^{(3)}_3(t)
 +8 \, (t-1)^{4}  \cdot   t^5 \cdot {\frac {d^{5}}{dt^{5}}}{C^{(3)}_3} (t)
\nonumber\\
&\,\, =\, \, \, \, - \frac {3 \,  ( t^2-1)\cdot  N^2\, ( 2N+1)^6} 
{(N+1)^3 } \cdot t^{N+3} \cdot B_0(N).
\label{c33ode}
\end{align}

From this differential equation we obtain the recursion relation for
the coefficients $c^{(3)}_{3;n}$ and the normalization constant
$A^{(3)}_3$ defined by the form (\ref{c02form}),
where by definition $c^{(3)}_{3;n}=\,0$ for $n\leq -1$
\begin{align}
&A^{(3)}_3 \cdot \{
8\,n \left( 2\,N-n \right)  \left( N-n \right)  \left( N+n \right) 
 \left( 3\,N-n \right)\cdot c^{(3)}_{3;n}\nonumber\\ 
&+4\, \left( 2\,N+1-2\,n \right)  \left( 2-7\,n+7\,N-{N}^{2}+4\,{n}^{4}-
12\,{N}^{3}-8\,{n}^{3}+24\,{N}^{3}n\right.\nonumber\\
&\left.-4\,{N}^{2}n+11\,{n}^{2}+4\,{N}^{2}
{n}^{2}-16\,N{n}^{3}+24\,N{n}^{2}-22\,Nn
\right)\cdot c^{(3)}_{3;n-1}\nonumber\\
&-16\, \left( N+1-n \right)  \left( 9-22\,n+22\,N+{N}^{2}+3\,{n}^{4}-18
\,{N}^{3}-12\,{n}^{3}+18\,{N}^{3}n\right.\nonumber\\
&\left.-6\,{N}^{2}n+23\,{n}^{2}+3\,{N}^{2}{
n}^{2}-12\,N{n}^{3}+36\,N{n}^{2}-46\,Nn \right)\cdot c^{(3)}_{3;n-2}\nonumber\\ 
&+4\, \left( 2\,N+3-2\,n \right)  \left( 32-69\,n+69\,N+7\,{N}^{2}+4\,{n
}^{4}-36\,{N}^{3}-24\,{n}^{3}\,+24\,{N}^{3}n\right.\nonumber\\
&\left.-12\,{N}^{2}\,n\,+59\,{n}^{2}+4
\,{N}^{2}{n}^{2}-16\,N\,{n}^{3}\, +72\,N{n}^{2}-118\,N\,n
\right)\cdot c^{(3)}_{3;n-3}\nonumber\\
&-8\, \left( n-2 \right)  \left( 2\,N+2-n \right)  \left( N+2-n
 \right)  \left( N-2+n \right)  \left( 3\,N+2-n \right)\cdot c^{(3)}_{3;n-4}\}
\nonumber\\ 
&=\, \, (\delta_{n,N}-\delta_{n,N+2})\cdot \frac{3(2N+1)^6}{(N+1)^3} \cdot B_0.
\label{recrelc33}
\end{align}

We note by sending $n\rightarrow 2N-n+2$ that $c^{(3)}_{3;n}$ and
$c^{(3)}_{3;2N-2-n}$ satisfy the same equation.

For $n=0$ (\ref{recrelc33}) is  identically zero for any
$c^{(3)}_{3;0}$ which we set equal to unity by convention.
For $0\leq n  \leq N-1$ the rhs of (\ref{recrelc33})  
vanishes and hence the
$c^{(3)}_{3;n}$ are identical with the  coefficients (\ref{c33nsec4}) 
of the solution (\ref{c33sol})
ot the homogeneous equation. 

For $n=N$ the coefficient of $c^{(3)}_{3;N}$ vanishes, and thus if there
were no inhomogeneous term, the coefficients $c^{(3)}_{3;n}$ for
$n=N-4,~N-3~,N-2~N-1$ would have to satisfy a non trivial constraint. This
constraint does not, in fact, hold and is the reason that the
homogeneous equation has a term $t^{N+3}\ln t$. However, with a
nonvanishing inhomogeneous term, the equation for $n=N$ determines the
normalization constant. 

For $n=N+1$ the equation (\ref{recrelc33}) reduces to
\begin{align}
&8\, \left( N+1 \right)  \left( N-1 \right)  \left( 2\,N+1 \right) 
 \left( 2\,N-1 \right)\cdot (c^{(3)}_{3;N+1}-c^{(3)}_{3,N-4})\nonumber\\
&-8\, \left( 2\,N+1 \right)  \left( 2\,N-1 \right)  \left( 2\,{N}^{2}-1
 \right)\cdot (c^{(3)}_{3;N}-c^{(3)}_{3;N-2})\,\,\,=\,\,\,\,0 
\end{align}
which will be satisfied by the palindromic property
\begin{equation}
c^{(3)}_{3;n}\,\,\,=\,\,\,\, c^{(3)}_{3;2N-2-n}
\end{equation}
with $n=\, N+1$ and $n=N$. Finally, the $c^{(3)}_{3;n}$ 
for $N\leq n\leq 2N-2$ are determined from the palindromy of 
(\ref{recrelc33}).

\section{Homomorphisms for $C_{0}^{(3)}(N;t)$ and $C_2^{(3)}(N;t)$}
\label{diffop}

The fifth order operator $M^{(3)}_0(N;t)$ in the direct sum
decomposition (\ref{dirsum3}) of $\Omega^{(3)}_0(N;t)$ has the homomorphism
(in terms of the operator $L_2(N)$)
\begin{eqnarray}
M^{(3)}_0(N)  \cdot J^{(3)}_0(N;t) \, \, = \, \, \, G^{(3)}_0(N;t) 
\cdot {\rm Sym}^4(L_2(N+1)), 
\end{eqnarray}
where the intertwinners $\,J^{(3)}_0(N;t)$ and $\, G^{(3)}_0(N;t)$ are:
\begin{eqnarray}
J^{(3)}_0(N;t) \, \, &= \, \, \,t^{N+1} \cdot (t-1) \cdot t \cdot 
\Bigl(D_t \, - \, {{d \ln(R^{A}_N) } \over {dt}}  \Bigr)
 \nonumber \\
 \, \, &= \, \, \, 
t^{N+1} \cdot \Bigl((t-1) \cdot t \cdot D_t \,
 -(2\, N \, + \, 2\, (N+1) ) \, \,  \Bigr),  \\
G^{(3)}_0(N;t) \, \, &= \, \, \,t^{N+1} \cdot (t-1) \cdot t \cdot
 \Bigl(D_t \, - \, {{d \ln(R^{B}_N) } \over {dt}}  \Bigr),  
\end{eqnarray}
 where
\begin{eqnarray}
R^{A}_N \, \, &= \, \, \,(t-1)^{2 \, (2\, N\, +1)} \cdot t^{-2 \, (N\, +1)}, \\
R^{B}_N \, \, &= \, \, \,  {{(t+1)\, (t-1)^{4\, N\, -3} } 
\over {t^{2\, N\, + 6} }} \cdot  P_N,\label{RAB} \\
P_N \, \, &= \, \, \, 
(4\,N+3)  \cdot (3\,N+2) \cdot ({t}^{2}+1)
\,  +2\, (20\,{N}^{2}\, +15\,N\, +2) \cdot  t
 \nonumber \\
  \, &= \,  \, (4\,N+3)  \cdot (3\,N+2)  \cdot (t+1) ^{2}\,  \\
&  \qquad +4\, \, ( 2\,  (2\,N+1) \, (N-1)\,  +N) \cdot t.
 \nonumber 
\end{eqnarray}

The homomorphism for $M^{(3)}_2(N;t)$ is

\begin{eqnarray}
&&M^{(3)}_2(N;t) \cdot J^{(3)}_2(N;t) \, \, = \, \, \, 
G^{(3)}_2(N;t) \cdot  {\rm Sym}^4(L_2(N+1)), 
\end{eqnarray}
where the intertwinners $\,J^{(3)}_2(N;t)$ and $\,G^{(3)}_2(N;t)$ are
\begin{eqnarray}
&& J^{(3)}_2(N;t) \, \, = \, \, \, t^{N+2} \cdot \Bigl( (t-1)\cdot t \cdot D_t 
\, +\, 2\, \, (N+1)\cdot  t \, +2\,N \Bigr), \\
&&G^{(3)}_2(N;t) \, \, = \, \, \, t^{N+2} \cdot (t-1)\cdot 
t \cdot \Bigl( D_t \, 
- \, \, {{d \ln \Bigl(R^{B}_N\Bigl(-(N+1)\Bigr) \Bigr) } \over {dt}}\Bigr),
\end{eqnarray}
where $\, R^{B}_N$ is {\em exactly} the $\, R^{B}_N$ in (\ref{RAB}). 

\section{Homomorphisms for $\Omega^{(4)}_4(N;t)$}
\label{more}

Many exact results have been obtained on the
intertwinners occurring in (\ref{m7}),
 (\ref{m51}), (\ref{m52}), (\ref{m3}) .
Let us display the simplest ones. 

For $J^{(4)}_0(N;t)$ we have
\begin{align}
J^{(4)}_0(2;t)\,\,&=\,\,\,\,\, t^2\cdot (t+1)\cdot (2t^2+t+2),
\label{j0}
\nonumber\\
J^{(4)}_0(3;t)\,\,&=\,\,\,\,\,t^2\cdot \left( t+1 \right)\cdot 
 \left( 64\,{t}^{4}+16\,{t}^{3}+99\,{t}^{2}+16\,t+64 \right),
\nonumber\\
J^{(4)}_0(4;t)\,\,&=\,\,\,\,\,t^2 \cdot\left( t+1 \right)\cdot 
 \left(
576\,{t}^{6}+96\,{t}^{5}+730\,{t}^{4}+425\,{t}^{3}+730\,{t}^{2
}+96\,t+576 \right),\nonumber\\
J^{(4)}_0(5;t)\,\,&=\,\,\,\,\,\ t^2\cdot\left( t+1 \right)\cdot 
 \left( 16384\,{t}^{8}+2048\,{t}^{7}+19264\,{t}^{6}\right.
\nonumber\\
&\qquad\qquad\left.+6608\,{t}^{5}+28861\,{t}^{4}+6608\,{t}^{3}+19264\,{t}^{2}+2048\,t
+16384 \right).
\end{align}

For $J^{(4)}_1(N;t)$ we have
\begin{align}
&2 \, t^3 \cdot J^{(4)}_1(2;t)\,\,  =\, \,\,  (t-1) \cdot  J^{(4)}_0(2;t) \cdot  D_t 
\, -2 \,t \cdot  (10t^4 \, +2t^3\,-5t\,-4), 
\nonumber \\
&64 \, t^4 \cdot J^{(4)}_1(3;t)\,\,  =\, \,\,  \left(t-1 \right) 
\cdot J^{(4)}_0(3;t)
      \cdot  D_t \nonumber \\
&  \qquad   -\, 2 \, t \cdot 
(448\,{t}^{6}+32\,{t}^{5}+95\,{t}^{4}-220\,{t}^{2}-112\,t-128),
\nonumber\\
&576 \cdot t^5 \cdot J^{(4)}_1(4;t) \,  =  \, \, (t-1) \cdot J^{(4)}_0(4;t) \cdot  
  D_t \nonumber\\
& -\, 2 \, t   \cdot  (5184
\,{t}^{8}+192\,{t}^{7}+406\,{t}^{6}+1148\,{t}^{5} 
 -2471\,{t}^{3}-1288\,{t}^{2}-864\,t-1152),\nonumber\\
& 16384 \cdot t^6 \cdot J^{(4)}_1(5;t)\, =\,\,\, \, (t-1)  \cdot 
 J^{(4)}_0(5;t) \cdot D_t \nonumber \\
& -\, 2  \, t \cdot 
(180224\,{t}^{10}+4096\,{t}^{9}+7488\,{t}^{8}
+15168\,{t}^{7}+41307\,{t}^{6} \nonumber \\
&-83454\,t^{4}-44112\,{t}^{3}-29952\,{t}^{2}-22528\,t-32768).
\end{align}
Finally, the simplest $J^{(4)}_2(N;t)$,  namely $J^{(4)}_2(2;t)$ reads:
\begin{align}
16\,t^6 \cdot J^{(4)}_2(2; t)\,\,&=\,\,\,\, 8 \, (t\, -1)^2 \cdot  J^{(4)}_0(2;t) \cdot D_t^2
\nonumber\\
&\qquad -\, t \cdot (t\, -1) \cdot (432t^4+80t^3-99t^2-240t-208) \cdot D_t
\nonumber\\
&\qquad + \, 3 \, (1040t^5-1176t^4-233t^3-100t^2+168t+256).
\end{align}

\section{Exact results for the $\, C^{(4)}_m$'s}
\label{C4j}

The $f^{(4)}_{N,N}(t)$'s have a new feature not previously seen.
The  inhomogeneous terms on the ODE's for $C^{(2)}_m(N;t)$ and 
$C^{(3)}_m(N;t)$ begin at $t^{N+a}$ where $a$ is 0,1 or 2 depending on
the values of $m$. Therefore to the order needed for the polynomial
solution the logarithms in the solution $u_2(N)$ never can contribute.
However, for $f^{(4)}_{N,N}(N;t)$ the order of the inhomogeneous terms
grows as $t^{2N}$ instead of $t^N$. Therefore, since logarithms occur
in $u_2(N)$ at order $t^{N+1}$ in order to find the
polynomial solution to the 20-th order inhomogeneous equation in terms
of the solutions of $u_2(N)$ and $u_1(N)$ we need to find linear
combinations of solutions of the terms in the direct sum decomposition
which cancel these logarithms.

This procedure for solving the inhomogeneous equations is too
cumbersome by itself to obtain explicit results
as was done for $f^{(2)}_{N,N}(t)$ and $f^{(3)}_{N,N}(t)$. However,
when the cancellation of logarithms is combined with the 
Wronskian cancellation method of section 5 it
is possible to conjecture results for $C^{(4)}_m(N;t)$ which have been
verified to satisfy the 20th order inhomogeneous equations through $N=10$:
\begin{align}
C^{(4)}_0 &\stackrel{2N+1}{=} \, \, -\bar{K}^{(4)}_0 \cdot 
\frac{u_2^4(N+1)}{t^4} \, \, \, 
- \bar{K}^{(4)}_0 \cdot \frac{4}{N} \cdot C^{(2)}_0 \cdot \frac{u^2(N+1)}{t^2}
  \nonumber\\
& + \frac{2}{3}\cdot \left[C^{(2)}_0 \cdot   \frac{u_2^2(N+1)}{t^2} \,\,
 -2\beta_N \cdot  C^{(2)}_0 \cdot  
\frac{u_2^2(N+1)\cdot u_2(N)}{t^2} \cdot  F_{N+1}\, \,  - C^{(2)}_1 \cdot 
\frac{u_2^3(N+1)}{t^3}\cdot F_{N+1}\right] \nonumber\\
& + \frac{N\lambda^2}{3} \cdot  \beta_N \cdot \frac{u_2^3(N+1)}{t^4}
 \cdot  t^{N+2} \cdot  F_{N+1}, 
\end{align}
\begin{align}
C^{(4)}_1 &\stackrel{2N+1}{=} \, \, 4 \cdot  \bar{K}^{(4)}_0 \cdot \beta_N 
\frac{u_2^3(N+1)u_2(N)}{t^3}  \nonumber\\
& +~ \bar{K}^{(4)}_0 \cdot \frac{4}{N}\cdot \left[2 \, \beta_N \cdot C^{(2)}_0 \cdot 
\frac{u_2(N+1)\cdot u_2(N)}{t}\, \,  - C^{(2)}_1 \cdot \frac{u_2^2(N+1)}{t^2} \right]
 \nonumber\\
& + \frac{2}{3} \cdot \left[6\, \beta_N^2 \cdot  C^{(2)}_0 \cdot
\frac{u_2(N+1)\cdot  u_2^2(N)}{t}\cdot  F_{N+1} \,  \,
+ 2\,C^{(2)}_1 \cdot \frac{u_2^3(N+1)}{t^3} \cdot F_N \,  \,
- 2\,C^{(2)}_2 \cdot \frac{u_2^3(N+1)}{t^3} \cdot F_{N+1}   \right]
 \nonumber\\
& - \frac{N\lambda^2}{3}
\left[\beta_N \cdot \frac{u_2^3(N+1)}{t^3} \cdot t^{N+1} \cdot F_N \,  \,
+ 3\beta_N^2 \cdot \frac{u_2^2(N+1) \cdot  u_2(N)}{t^3}\cdot t^{N+2} \cdot  F_{N+1} \right],
\end{align}
\begin{align}
C^{(4)}_2 &\stackrel{2N+1}{=}\,\, - 6\, \bar{K}^{(4)}_0
\cdot  \beta^2_N \cdot \frac{u_2^2(N+1)\cdot u^2_2(N)}{t^2} \nonumber\\
& -~\left(\frac{4}{N}\cdot \bar{K}^{(4)}_0 \, +2\right) 
\cdot \left[\beta_N^2 \cdot C^{(2)}_0  \cdot
u_2^2(N)\,\,  - 2 \beta_N \cdot  C^{(2)}_1 \frac{u_2(N+1)\cdot  u_2(N)}{t} \,
+ C^{(2)}_2 \frac{u_2^2(N+1)}{t^2}\right] \nonumber\\
& + \frac{2}{3}\cdot \left[-6\beta_N^3 \cdot C^{(2)}_0 \cdot u_2^3(N)\cdot F_{N+1} \,
-9\, \beta_N \cdot C^{(2)}_1 \cdot \frac{u_2(N+1)\cdot u_2(N)}{t} \, \,
+ 6\, C^{(2)}_2 \cdot \frac{u_2^3(N+1)}{t^3} \cdot F_N   \right] \nonumber\\
& + \frac{N\lambda^2}{3} \cdot \left[3 \beta_N^2 \cdot 
\frac{u_2^2(N+1) \cdot u_2(N)}{t^2} \cdot t^{N+1} \cdot  F_N \, \, +3 \cdot \beta_N^3 \cdot 
\frac{u_2(N+1) \cdot  u_2^2(N)}{t^2} \cdot  t^{N+2} \cdot F_{N+1} \right],
\end{align}
\begin{align}
C^{(4)}_3 &\stackrel{2N+2}{=} \, \, \, \bar{K}^{(4)}_04 \cdot \beta^3_N \cdot 
\frac{u_2(N+1)\cdot u^3_2(N)}{t} \nonumber\\
& -~ \bar{K}^{(4)}_0 \cdot \frac{4}{N} \cdot \left[\beta_N^2 \cdot  C^{(2)}_1 \cdot  u_2^2(N)  \, \,
- 2\, \beta_N \cdot  C^{(2)}_2 \cdot \frac{u_2(N+1)\, u_2(N)}{t}\right]
 \nonumber\\
& + \frac{2}{3} \cdot \left[2\, \beta_N^3 \cdot C^{(2)}_0 \cdot u_2^3(N) \cdot F_N \, \,
- 2\, \beta_N^3 \cdot C^{(2)}_1 \cdot u_2^3(N) \cdot F_{N+1}  \,\,
- 6\, \beta_N \cdot C^{(2)}_2 \cdot \frac{u_2^2(N+1)  \cdot  u_2(N)}{t^2 \cdot }\cdot F_N   \right]
 \nonumber\\
& - \frac{N\lambda^2}{3} \cdot 
\left[3\, \beta_N^3 \cdot \frac{u_2(N+1) \cdot  u_2^2(N)}{t} \cdot 
t^{N+1} \cdot F_N \, \, + \beta_N^4 \cdot \frac{u_2^3(N)}{t}\, t^{N+2} \cdot F_{N+1}  \right],
\end{align}
\begin{align}
C^{(4)}_4 &\stackrel{2N+3}{=}\,\, -\bar{K}^{(4)}_0 \cdot \beta^4_N \cdot  u^4_2(N) \, \, \,
-\bar{K}^{(4)}_0 \cdot \frac{4}{N} \cdot \beta_N^2 \cdot C^{(2)}_2 \cdot u_2^2(N)
 \nonumber\\
& + \frac{2}{3}\cdot \left[\beta_N^3 \cdot C^{(2)}_1\cdot u_2^3(N)\cdot F_N \,\,  
+ 2\, \beta_N^2\cdot C^{(2)}_2\cdot \frac{u_2(N+1)\cdot  u_2^2(N)}{t}\cdot F_N \, \, 
+ \beta_N^2 \cdot C^{(2)}_2 \cdot   u_2^2(N)  \right] \nonumber\\
& + \frac{N\, \lambda^2}{3} \cdot  \beta_N^4 \cdot  u_2^3(N)\cdot  t^{N+1} \cdot F_N
\end{align}
In order to construct the full $C^{(4)}_m$, the expressions above are 
series expanded up to the order of palindromy, with palindromy
determining the rest of the terms. The palindromy points of the 
$C^{(4)}_m$ are given as follows: $m=0:2N+1$, $m=1:2N+1$, $m=2:2N+2$, 
$m=3:2N+2$, $m=4:2N+3$. Therefore, the expressions above give all 
terms to all $C^{(4)}_m$ except for the middle term of $C^{(4)}_2$ 
at order $2N+2$, which is determined such that all terms in 
$f^{(4)}_N$ cancel up to and including $2N+3$.

Note that while these $C^{(4)}_m$ guarantee that all terms will 
vanish up to and including $2N+3$, it is not obvious that the 
expansion at order $2N+4$ will match the expansion 
of $f^{(4)}_N$, even though it is the case.

\end{document}